\documentclass[aps,prd,12pt, a4paper, nofootinbib,superscriptaddress, notitlepage, floatfix]{revtex4-2}
 \pdfoutput=1
\usepackage[usenames,dvipsnames]{color}  
\usepackage{graphicx}
\usepackage{setspace}
\usepackage{subfigure}
\usepackage{makecell}
\usepackage{amsmath}
\usepackage{amssymb}
\usepackage[colorlinks=true,citecolor=blue,urlcolor=blue, pdfborder={0 0 0}]{hyperref}
\usepackage[normalem]{ulem}
\usepackage{xcolor}
\usepackage{multirow}

\makeatletter
\def\p@subsection{}
\makeatother

\definecolor{darkred}{rgb}{0.6,0,0}

\definecolor{linkcolor}{rgb}{0,0,0.5}

\def\gsim{\raise0.3ex\hbox{$\;>$\kern-0.75em\raise-1.1ex\hbox{$\sim\;$}}}
\def\lsim{\raise0.3ex\hbox{$\;<$\kern-0.75em\raise-1.1ex\hbox{$\sim\;$}}}

\def\beqn#1{\begin{equation}\label{#1}}
\def\eeqn{\end{equation}}

\def\beqa#1{\begin{eqnarray}\label{#1}}
\def\eeqa{\end{eqnarray}}

\def\Z2{$\mathcal{Z_2}$}

\newcommand {\ignore}[1]{}

\def\321{$\mathrm{SU(3) \otimes SU(2) \otimes U(1)}$ }

\def\bl#1{\textcolor{blue}{#1}}

\newcommand{\AddrIACS}{School of Physical Sciences, Indian Association for Cultivation of Science,\\
2A $\&$ 2B Raja S C Mullick Road, Kolkata 700032, India}

\newcommand{\AddrIOP}{
Institute of Physics, Sachivalaya Marg, Bhubaneswar, 751 005, India ; Homi Bhabha National Institute, Training School Complex, Anushakti Nagar, Mumbai 400 094,
India}
 
 \newcommand{\AddrIISERB}{Department of Physics,
 Indian Institute of Science Education and Research - Bhopal \\
 Bhopal Bypass Road, Bhauri, Bhopal, India}

\begin{document}
\title {The $N_{\rm eff}$ at CMB challenges $U(1)_X$ light gauge boson scenarios}
\author{Dilip Kumar Ghosh}\email{dilipghoshjal@gmail.com}
\affiliation{\AddrIACS}
\author{Purusottam Ghosh}\email{pghoshiitg@gmail.com}
\affiliation{\AddrIOP}
\author{Sk Jeesun}\email{skjeesun48@gmail.com}
\affiliation{\AddrIACS}
\author{Rahul Srivastava}\email{rahul@iiserb.ac.in}
\affiliation{\AddrIISERB}


\begin{abstract}
 \vspace{0.5cm} 
 \begin{center}
  \large{Abstract} 
\end{center}
The relativistic degrees of freedom ($N_{\rm eff}$) is one of the crucial cosmological parameters. The precise measurement of $N_{\rm eff}$ at the time of cosmic microwave background formation, by Planck 2018 can be used to understand the new fundamental interactions, in particular involving light mediators. 
Presence of any new particle with sufficient energy density and sizeable interactions with Standard Model particles at the temperature around $\sim$ MeV can significantly alter the neutrino decoupling and hence $N_{\rm eff}$.
Thus the bound on $N_{\rm eff}$ can place stringent constraints on various beyond Standard Model paradigms involving light particles.
$U(1)_X$ models are among such scenarios and are widely studied in several aspects. 
In this work, we consider several popular $U(1)_X$ models with light $Z'$ boson like  $U(1)_{B-L}$, $U(1)_{B - 3L_i}$, $U(1)_{B_i - 3 L_j}$, $U(1)_{L_i - L_j}$; $i,j =1,2,3$ being the flavour indices and study their impact on $N_{\rm eff}$.
We also examine the constraints from ground based experiments like Xenon1T, Borexino, trident, etc.
Our analysis shows that for light mass $M_{Z'} \lesssim \mathcal{O} (\rm{MeV})$ the $N_{\rm eff}$  provides the most stringent constraints on the $Z'$ mass and coupling, far exceeding the existing constraints from other experiments.  

\end{abstract}
\maketitle


\section{INTRODUCTION}
\label{sec:intro}
 The Cosmic Microwave Background (CMB) radiations play a crucial role in elucidating the 
dynamics of the early universe as well as shaping our present understanding of the large-scale structure 
of the universe. 
Observations of the CMB, including temperature anisotropies, polarization patterns, matter-energy distribution, reionization phenomena in the early universe, are consistent with both the standard $\Lambda$CDM cosmology and the Standard Model of particle physics (SM) \cite{Dodelson:2003ft}.
One of the interesting cosmological parameters $N_{\rm eff}$, which represents the number of relativistic degrees of freedom at the CMB scale at temperature \footnote{{ The temperature of the universe is denoted as $T$ and generally defined as photon bath temperature ($T_\gamma$)}} around $T\sim$ eV, serves an important role in understanding the dynamics of the thermal history during the epoch $T\sim$ MeV to eV. At high temperature of the universe, photons and neutrinos shared the same temperature. 
As the universe cooled down ($T \lesssim 2$ MeV) and the interaction rate fell below the Hubble expansion rate ($H$), neutrinos decoupled from photons, resulting in the formation of two separate baths i.e. neutrino bath and photon bath (combined with electrons). 
At temperature $T\lesssim m_e$, the two baths evolve with different temperature \cite{Dodelson:2003ft}. 
The parameter $N_{\rm eff}$ is then parameterized by the ratio of energy densities in the neutrino and photon baths. In the particle content of the SM, one can expect the neutrino decoupling event to take place around $T \sim 2$ MeV, resulting in $N_{\rm eff}^{\rm SM}= 3.046$ at the CMB \cite{Mangano:2005cc,Grohs:2015tfy}. 
Note the slight deviation in $N_{\rm eff}$ from the expected three light neutrino degrees of freedom in SM ($3~\nu_L$) is due to non-instantaneous neutrino decoupling, finite temperature QED corrections, and neutrino flavor oscillations  \cite{Mangano:2005cc,Grohs:2015tfy}. 
\begin{figure}[!tbh]
    \centering
    \includegraphics[scale=0.55]{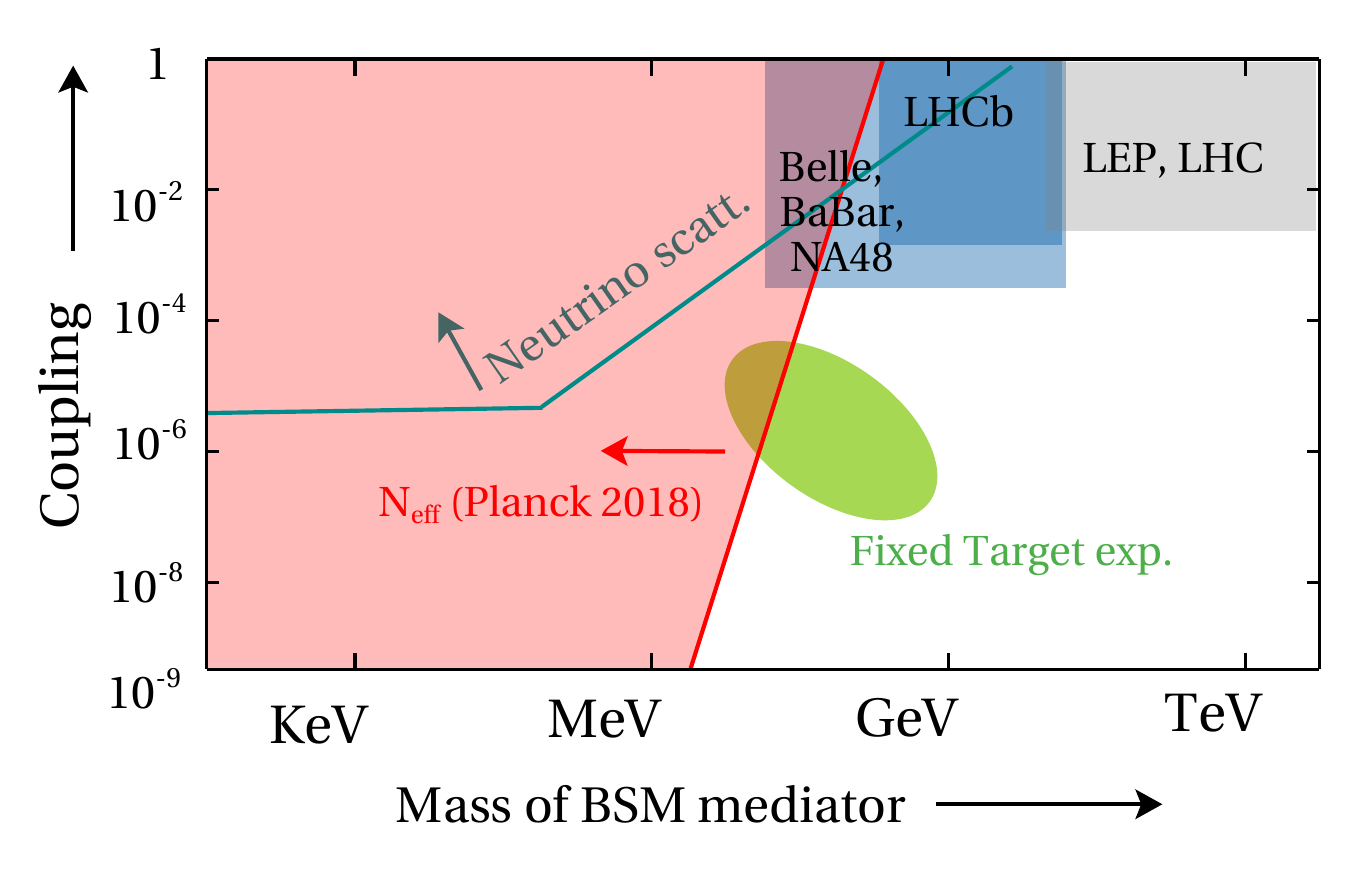}
    \caption{Schematic diagram of the constraint from $N_{\rm eff}$ along with existing constraints for light mediator models.
    This schematic diagram assumes universal coupling of the mediator to both quarks and the leptons. Note that additional bounds may apply as well as some of these bounds can be relaxed depending on the specific details of different models. }
    \label{fig:cartoon}
\end{figure}
Observations of the CMB by the Planck satellite can also measure the parameter $N_{\text{eff}}$ \footnote{Measurements of baryon acoustic oscillations (BAO), large-scale structure (LSS) datasets, and other cosmological probes contribute to determining the value of $N_{\rm eff}$ \cite{Planck:2015fie,Planck:2018vyg}.}. The latest Planck 2018 data provides a precise measurement of $N_{\rm eff}$ at the time of the CMB to be $2.99^{+0.34}_{-0.33}$ with $95\%$ confidence level (CL), under the assumption of standard $\Lambda$CDM cosmology \cite{Planck:2018vyg}.
The upper limit of the Planck data suggests an additional contribution (apart from SM) to $N_{\rm eff}$, extending up to $0.28$ ($\Delta N_{\rm eff}=N_{\rm eff}^{\rm obs}-N_{\rm eff}^{\rm SM}$), providing the hint for BSM physics. 

The presence of beyond standard model (BSM) light mediators, interacting with either the photon or neutrino bath at temperatures relevant to neutrino decoupling, introduces additional contributions to the radiation energy density, leading to an increase in $N_{\rm eff}$ \cite{Escudero:2018mvt,Escudero:2019gzq,Esseili:2023ldf,Li:2023puz} \footnote{Note additional contributions to $N_{\rm eff}$ may also come from various other sources like extra radiation \cite{Abazajian:2019oqj, Poulin:2018cxd, Ghosh:2022fws, Ghosh:2023ocl}, models involving early dark energy \cite{Poulin:2018cxd, Poulin:2018dzj}, and relativistic decaying dark matter \cite{Bringmann:2018jpr}.}. The observations of $N_{\rm eff}$ impose constraints on the BSM light mediators, as presented by the red shaded region in the cartoon Fig.\ref{fig:cartoon}. In certain mass regions of these mediators, $N_{\rm eff}$ turns out to be a severe constraint compared to other existing bounds. 
$U(1)_X$ models are among such widely studied BSM paradigms with light mediators and their phenomenological consequences have been explored in several contexts\footnote{See for example Ref. \cite{Bonilla:2017lsq,Bauer:2018onh,Coloma:2020gfv,Coloma:2022umy,AtzoriCorona:2022moj,Majumdar:2021vdw,DeRomeri:2024dbv,Chakraborty:2021apc,Demirci:2023tui} and the references therein.}.
In this work, we consider  light gauge boson ($Z^\prime$) originating from different types of $U(1)_X$ gauge extended scenarios, which naturally involve interactions with neutrinos and electron-positron pairs (either at the tree level or loop-induced). 
The BSM interactions in the presence of light $Z'$ can potentially modify the late-time dynamics between the photon and neutrino baths, prolonging the process of neutrino decoupling and increasing the $N_{\rm eff}$ value. As a result, the upper limit on $N_{\rm eff}$ ($3.33$ with $95\%$ CL) derived from the CMB can be utilized to constrain both the mass of the light gauge boson ($M_{Z^\prime}$) and its corresponding gauge coupling ($g_X$).

In the context of $N_{\rm eff}$, various types of gauged $U(1)_X$ scenarios have been explored in the existing literature. These scenarios aim to explain the excess of $N_{\rm eff}$ observed in the CMB, involving $U(1)_{L_\mu-L_\tau}$ \cite{Escudero:2019gzq}, $U(1)_{B-L}$ \cite{Esseili:2023ldf} or, generic $U(1)_X$ model \cite{Ghosh:2023ilw}. Here $B,L$ signify Baryon and Lepton numbers respectively \footnote{Usually the $X$ in $U(1)_X$ models are named after the $U(1)$ charges of quark (fundamental constitutents of baryons) or lepton fields e.g. for $U(1)_{B-L}$ every quark (lepton) acquires charges equivalent to $B-L=1/3(-1)$ under the  new $U(1)$ symmetry.}.
A comprehensive discussion on $N_{\rm eff}$ within the generic $U(1)_X$ scenario is presented in  Ref.   \cite{Ghosh:2023ilw}, categorizing $U(1)_X$ models into two classes: $(i)$ those with tree-level coupling of $Z^\prime$ with electron and $(ii)$ those without tree-level coupling of $Z^\prime$ with electron. 
The study also highlights that the varying $U(1)_X$ charges for $e, \nu_{e,\mu,\tau}$, corresponding to various gauged extended scenarios, notably impact the value of $N_{\rm eff}$. This work is an extension of  Ref.  \cite{Ghosh:2023ilw}, where we consider a variety of $U(1)_X$ models, such as $U(1)_{B-L}$, $U(1)_{L_i-L_j} (i\neq j)$, $U(1)_{B-3L_i}$, and $U(1)_{B_i-L_j}$ etc., and study their impact on $N_{\rm eff}$, assuming the scenario where $Z^\prime$ was initially ($T> M_{Z^\prime}$) in thermal bath
(which requires $g_X \gtrsim \mathcal{O}(10^{-9})$). The primary objective of this work is to identify the exclusion regions in the $M_{Z^\prime} - g_X$ plane for each gauged extended model, guided by the upper bound on $N_{\rm eff}$ derived from CMB observations \cite{Planck:2018vyg}.

The gauged extended models are also motivated by their potential detectability across a broad mass range, from eV to TeV, as sketched in Fig.\ref{fig:cartoon}. For gauge boson $Z^\prime$ with TeV or sub TeV masses, the most severe constraints are derived from collider experiments like LHC \cite{CMS:2016cfx,ATLAS:2019erb,Das:2016zue,Accomando:2017qcs} and LEP \cite{Essig:2009nc}. Whereas, in the mass range spanning from a few MeV to a few GeV, experiments carried out by LHCb \cite{LHCb:2017trq}, Belle\cite{Inguglia:2016acz}, BaBar \cite{BaBar:2014zli}, and fixed target experiments \cite{Bauer:2018onh,Bross:1989mp} are notably more effective in providing constraints. In the sub GeV mass region, with  $g_X \gtrsim \mathcal{O}(10^{-6})$, the low energy neutrino scattering experiments (neutrino electron scattering \cite{Coloma:2022umy}, neutrino-nucleus scattering \cite{COHERENT:2021xmm} etc.), provide most stringent constraints. However, in the mass range where $M_{Z^\prime} \lesssim \mathcal{O}$(MeV) and $g_X \lesssim \mathcal{O}(10^{-6})$, direct searches become less sensitive. 
 Furthermore, as we will show, CMB observations in terms of $N_{\rm eff}$ play a significant role in constraining the parameter space for $M_{Z^\prime} \lesssim \mathcal{O} ({\rm MeV})$.
 In this work, we identify the CMB exclusion region with $95\%$ CL for each $U(1)_X$ scenario and compare it to the other aforementioned existing constraints in the $M_{Z^\prime} - g_X$ plane.
We highlight that in certain regions of $M_{Z^\prime}$ and coupling $g_X$, observations of $N_{\rm eff}$ in the CMB provide more stringent constraints compared to other observations. 
We present a schematic diagram in Fig.\ref{fig:cartoon}, where various existing constraints are summarised by different color patches based on the mass and coupling. Our key focus in this study is the red-shaded region, which represents the exclusion region derived from $N_{\rm eff}$ observations in the CMB.

The paper is organized in the following way. In section \ref{sec:neff}, we provide a brief overview of the evolution of $N_{\rm{eff}}$ in the presence of a light $Z^\prime$, with a focus on its dependence on the gauge extension.
Then we present the exclusion regions in the $M_{Z^\prime}-g_X$ plane for various $U(1)_X$ models, determined by the upper bound on $N_{\rm{eff}}$ derived from CMB observations in section \ref{sec:examples}. Finally, in section \ref{sec:conclusions} we summarize our results and conclusions.

\section{Evaluation of $N_{\rm eff}$ in light $Z'$ Models}
\label{sec:neff}
Before going to the $N_{\rm eff}$ analysis of light $Z'$ emerging from various $U(1)_X$ symmetries we 
present a brief and generic overview of the effective $Z'$ models in this section. 
For a generic $U(1)_X$ model, we assume that SM quark doublets ($Q_i$) and singlets ($u_i, d_i$), as well as lepton doublets ($L_i$) and singlets ($\ell_i$) are charged under this $U(1)_X$ symmetry\footnote{Note that for now we have left the $U(1)_X$ charges of SM particles as free parameters, whose values depend on the details of the $U(1)_X$ symmetry, see Table~\ref{tab:all}.}. 
We also need a SM singlet scalar $\sigma$ carrying $U(1)_X$ charge to generate 
the mass of $Z'$ ($M_{Z'}$) by acquiring a vacuum expectation value (VEV) to break the $U(1)_X$ symmetry. 
As pointed out in  Ref. \cite{Ghosh:2023ilw} that to affect the $\nu_L$ decoupling $Z'$ has to be in the MeV scale.
To keep $M_{Z^\prime} $ very light compared to the SM $Z$ gauge boson ($M^2_{Z^\prime} \ll M^2_{Z}$), we consider $U(1)_X$ 
charge of the SM Higgs doublet $\mathbb{X}_\Phi = 0$ and hence $M^2_{Z^\prime} $ does not get any contribution from the SM vev.
The charge assignments are shown in Table~\ref{tab:all}.

Note that to generate the SM quark and charged lepton masses, we take the following $U(1)_X$ charge assignments: $\mathbb{X}_{Q_i} = \mathbb{X}_{u_i} = \mathbb{X}_{d_i}$ and $\mathbb{X}_{L_i} = \mathbb{X}_{\ell_i}$.
It is also worth highlighting that although the $U(1)_X$ charges of quark doublet should be the same as that of the up and down quark singlets, the charges  may differ  across different generations i.e. 
$\mathbb{X}_{Q_1} \neq \mathbb{X}_{Q_2} \neq \mathbb{X}_{Q_3}$ as shown in Table~\ref{tab:all}. 
To simplify our notation throughout this paper we will denote $\mathbb{X}_{L_i} = \mathbb{X}_{\ell_i}=\mathbb{X}_i$.

\renewcommand{\arraystretch}{1.4}
 \begin{table}[!tbh]
\begin{center}
\begin{tabular}{| c | c | c | c | c |}
  \hline
~~~Models~~~  &$~~~\mathbb{X}_{Q_i}(\mathbb{X}_{u_i}=\mathbb{X}_{d_i})~~~$& ~~$\mathbb{X}_{L_1}$~~  & ~~$\mathbb{X}_{L_2}$~~
& ~~$\mathbb{X}_{L_3}$~~    \\
\hline \hline
$\mathbf {B-L}$  & $(1/3,~1/3,~1/3)$    & $-1$       & $-1$      & $-1$ \\
\hline
$\mathbf {B-3L_e}$  & $(1/3,~1/3,~1/3)$    & $-3$       & $0$      & $0$  \\

${B-3L_\mu}$  & $(1/3,~1/3,~1/3)$    & $0$       & $-3$      & $0$  \\

${B-3L_\tau}$  & $(1/3,~1/3,~1/3)$    & $0$       & $0$      & $-3$  \\
\hline
$\mathbf {L_e-L_\mu}$  & $(0,0,0)$                 & $1$       & $-1$      & $0$  \\ 
$\mathbf {L_e-L_\tau}$  & $(0,0,0)$                 & $1$       & $0$      & $-1$ \\
${L_\mu-L_\tau}$  & $(0,0,0)$                 & $0$       & $1$      & $-1$ \\
\hline
$\mathbf {B_1-3L_e}$  & $(1,0,0)$    & $-3$       & $0$      & $0$  \\
$\mathbf {B_2-3L_e}$  & $(0,1,0)$    & $-3$       & $0$      & $0$ \\
$\mathbf {B_3-3L_e}$  & $(0,1,0)$    & $-3$       & $0$      & $0$ \\
\hline 
${B_1-3L_\mu}$  & $(1,0,0)$    & $0$       & $-3$      & $0$  \\
${B_2-3L_\mu}$  & $(0,1,0)$    & $0$       & $-3$      & $0$ \\ 
${B_3-3L_\mu}$  & $(0,1,0)$    & $0$       & $-3$      & $0$ \\ 
\hline 
${B_1-3L_\tau}$  & $(1,0,0)$    & $0$       & $0$      & $-3$  \\
${B_2-3L_\tau}$  & $(0,1,0)$    & $0$       & $0$      & $-3$ \\ 
${B_3-3L_\tau}$  & $(0,1,0)$    & $0$       & $0$      & $-3$ \\ 
\hline 
  \end{tabular}
\end{center}
\caption{ $U(1)_X$ gauge charge assignments of SM particles for different models. To ease our notation throughout this paper we denote $\mathbb{X}_{L_i} = \mathbb{X}_{\ell_i}$. The models with tree level $Z'e^+e^-$ vertex ($\mathbb{X}_{1}\neq0$)  are written in bold front. The charges of $\nu_{R_i}$ are fixed by anomaly cancellation condition and the charge of the scalar singlet $\sigma$, needs to give mass to $Z'$, depending on the model and details of neutrino mass generation. For the rest of this paper we will denote $\mathbb{X}_{L_i} = \mathbb{X}_{\ell_i}=\mathbb{X}_i$. As mentioned earlier the models are named after the $U(1)_X$ charges of quark (fundamental constituents of baryons) and lepton fields. }
  \label{tab:all}
\end{table}
\renewcommand{\arraystretch}{1.0}
Besides the particle mentioned in Table~\ref{tab:all}, we also need right-handed neutrino (RHN) $\nu_R$ for anomaly cancellation and light neutrino mass generation \cite{Ma:2014qra,Ma:2015mjd,Bonilla:2017lsq}.
In this work, we consider only Majorana type mass models where, $\nu_{R}$ are also too heavy to affect $\nu_L$ decoupling \cite{Mohapatra:1979ia,Schechter:1980gr} and we ignore their contribution\footnote{In Dirac-type mass models, $\nu_{R}$ are relativistic at MeV temperature and can significantly alter $N_{\rm eff}$ \cite{Abazajian:2019oqj,Luo:2020sho,Luo:2020fdt}.}.
Apart from the minimal particle content of  Table~\ref{tab:all}, most of the popular $U(1)_X$ models studied in literature may also contain additional BSM particles which are typically much heavier than MeV scale and will not take part in our analysis \cite{Bonilla:2017lsq}.
%
%
\begin{figure}[!tbh]
    \centering
    \includegraphics[scale=0.6]{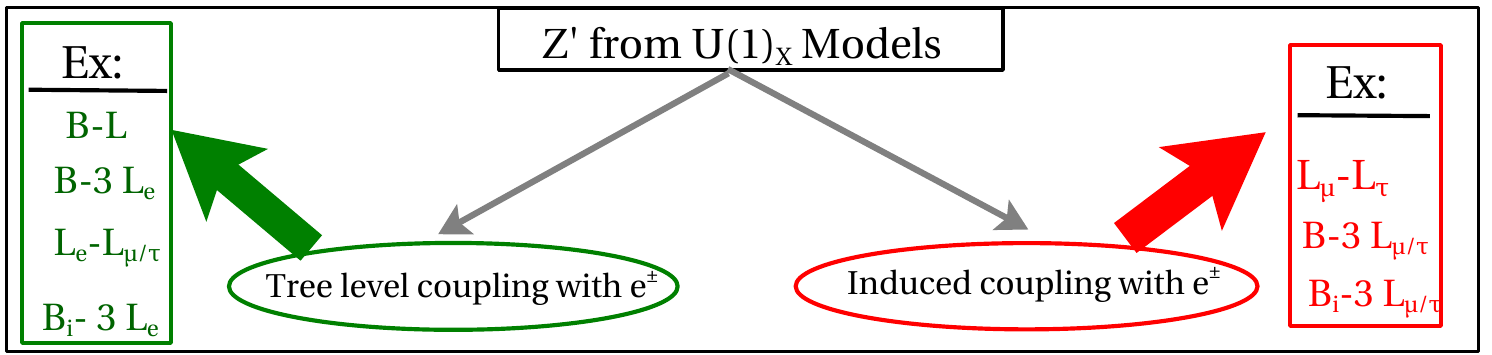}
    \caption{The $U(1)_X$ models can be broadly classified into two categories: (a) those where $Z'$ couples at tree level with electrons ($\mathbb{X}_{1}\neq0$), encircled in green color and (b) those where $Z'$ has no tree level coupling with electrons ($\mathbb{X}_{1}=0$), encircled in red color. See text for more details. We denote $\mathbb{X}_{L_i}$ as $\mathbb{X}_i$.}
    \label{fig:class}
\end{figure}

The $\nu_L$ decoupling is governed by only the weak interaction processes in SM scenario and usually takes place at temperature around $T\sim $ MeV temperature \cite{Escudero:2018mvt}. 
The only particles relevant at that temperature are $e^\pm, ~\gamma$ and $\nu_i~(i\equiv e,\mu,\tau)$. 
The scenario changes drastically in the presence of any light (mass $\sim \mathcal{O}(1)$ MeV) BSM particles interacting with either of these three sectors \cite{Escudero:2019gzq,Ghosh:2023ilw}.
Thus one can easily anticipate that light $U(1)_X$ gauge boson $Z'$ (with mass, $M_{Z'}\sim \mathcal{O}(1)$ MeV) interacting with $\nu_i$ and $e^\pm$  will significantly alter the $N_{\rm eff}$ at the time of CMB.
In this work we will assume all $3$ $\nu_i$ acquire the same temperature ($T_{\nu_L}$) and consider the case where $Z'$ was initially in a thermal bath ($g_X\gtrsim 10^{-9}$) \cite{Ghosh:2023ilw}.
$N_{\rm eff}$ is parametrised in terms of the temperature ratios of $\nu_i$ sector and $\gamma$ bath \cite{Dodelson:2003ft},
\begin{eqnarray}
    N_{\rm eff}= \frac{8}{7} \left(\frac{11}{4} \right)^{4/3} \left(\frac{\rho_{\nu_L}(T_{\nu_L})}{\rho_\gamma(T_\gamma)} \right)=3\times\frac{8}{7} \left(\frac{11}{4} \right)^{4/3} \left(\frac{ T_{\nu_L}}{T_\gamma} \right)^4
    \label{eq:neff}
\end{eqnarray}
$N_{\rm eff}$ at the time of CMB is determined by computing the temperature ratio $ \frac{T_{\nu_L}}{T_{\gamma}}$ at $T_\gamma=T_{\rm CMB}$.
Note that, throughout this paper, whenever we say $N_{\rm eff}$, we refer $N_{\rm eff}$ at the time of CMB.
To evaluate $T_{\nu_L}/T_{\gamma}$ we solve the following coupled equations \footnote{Further details can be found in Appendix A-C of  Ref.  \cite{Ghosh:2023ilw}.},
\begin{eqnarray}
\frac{dT_{\nu_L}}{dt} &=& -  \left( 4 H  \rho_{\nu_L}   -  \frac{ \delta \rho_{\nu_{L} \to e^{\pm}}}{\delta t}+  \frac{ \delta \rho_{Z'\to\nu_{L}}}{\delta t}    \right) \left(   \frac{\partial \rho_{\nu_{L}} }{ \partial T_{\nu_{L}} }    \right)^{-1} \label{eq:nux} \\
\frac{dT_{Z'}}{dt} &=& -  \left( 3 H \left( \rho_{Z'} + P_{Z'}\right) -  \frac{ \delta \rho_{Z'\to\nu_{L}}}{\delta t} -  \frac{ \delta \rho_{Z'\to e^{\pm}}}{\delta t}    \right) \left( \frac{\partial \rho_{Z'}}{ \partial T_{Z'}  }  \right)^{-1} 
\label{eq:Tnu_mu-tau} \\
\frac{dT_{\gamma}}{dt} & =&- \left(  4 H \rho_{\gamma} + 3 H \left( \rho_{e} + P_{e}\right) + \frac{ \delta \rho_{\nu_{L}\to e^{\pm}}}{\delta t} +  \frac{ \delta \rho_{Z'\to e^{\pm}}}{\delta t}  \right)\left(    \frac{\partial \rho_{\gamma}}{\partial T_\gamma} + \frac{\partial \rho_e}{\partial T_\gamma}       \right)^{-1}
   \label{eq:Tgamma},
\end{eqnarray}

where, $\rho_{r}, P_r$ and $T_r$ stand for the energy density, pressure density, and temperature of species $r$.
On the other hand, $\frac{\delta\rho_{a \to b}}{\delta t}$ signify the energy transfer rate from bath $a$ to $b$, deduced  by integrating the collision terms \cite{Escudero:2019gzq,Ghosh:2023ilw}.
As mentioned earlier, assuming same temperature between 3 $\nu_i$ sectors, we denote $\rho_{\nu_L}=\sum_{i=e,\mu,\tau}\rho_{\nu_i}$,
$\frac{\delta \rho_{\nu_L\to e}}{\delta t}=\sum_{i=e,\mu,\tau}\frac{\delta \rho_{\nu_i\to e}}{\delta t}$ and $\frac{ \delta \rho_{Z'\to \nu_L}}{\delta t}=\sum_{i=e,\mu,\tau} \frac{ \delta \rho_{Z'\to \nu_i}}{\delta t}$.
Throughout the paper, by $\nu_L$ we refer all $3$ generation of $\nu_{i}~(i\equiv e,\mu,\tau)$ as a whole.
We do not discuss the energy transfer rates here as they are already elaborated in detail in  Ref.  \cite{Ghosh:2023ilw}.

It is worth mentioning that the $ \nu_i\Bar{\nu_i}Z^\prime $ coupling and the $e^+e^- Z^\prime $ coupling are the most crucial parameters in the BSM processes that affect the $N_{\rm eff}$. However, interactions involving heavier charged leptons and quarks are insignificant here, as their energy densities are suppressed at temperatures nearing $\sim 1$ MeV.
A detailed discussion with the numerical estimation of $N_{\rm eff}$ for popular $U(1)_X$ models is made in Appendix \ref{app:neff_all} which we summarise here.
When $Z'$ has coupling to both $\nu_L$ and $e^\pm$ ($\mathbb{X}_{1}\neq 0$) , both 
the following BSM processes affect $\nu_L$ decoupling:
(i) $Z'$ decaying to both $e^{+}e^{-}$ and $ \nu_i \Bar{\nu_i}$ and 
(ii) scattering process $\nu_{i}\Bar{\nu_{i}} \to e^+e^-$ mediated by $Z'$. 
Any increase in the effective coupling ($\mathbb{X}_1 g_X$) dictating these BSM processes enhances the BSM contribution,
thus leading to an increment in $N_{\rm eff}$.
On the other hand in the absence of tree level $Z'e^+e^-$ vertex $( \mathbb{X}_1 =0)$, all the number density of 
$Z'$ finally gets diluted to $\nu_L$ bath (with $100\%$ branching ratio) irrespective of the coupling strength ($\mathbb{X}_{2/3}g_X$).
Hence for such models ($ \mathbb{X}_1 =0$), the final $N_{\rm eff}$ becomes independent of effective couplings $\mathbb{X}_{2,3}g_X$.
Thus following the above argument (also see Appendix \ref{app:neff_all})  the light $Z'$ models can be broadly classified into two cases: $\mathbb{X}_{1}\neq0$ and $\mathbb{X}_{1}= 0$, depending on whether $Z'$ has tree level coupling with electron or not \cite{Ghosh:2023ilw} (see Fig.\ref{fig:class}).

 It is worth highlighting that even in the absence of tree level $Z' e^+e^-$ interaction ($\mathbb{X}_1=0$), induced coupling with $e^\pm$ may exist which varies depending on the specific model \cite{Amaral:2020tga}. 
We take the following effective $Z^\prime e^+e^-$ (induced) interaction Lagrangian:
 \begin{equation}
     \mathcal{L}_{int}= (\epsilon e) \Bar{e} \gamma^\mu e Z'_\mu,
     \label{eq:mixL}
 \end{equation}  
 Note that such induced coupling in a given model can arise in different ways. For example it may arise from (i) tree level kinetic mixing of $U(1)_X$ and $U(1)_Y$ gauge bosons (ii) loop induced effective coupling (iii) mixing among charged leptons.
 We do not delve into the details of induced coupling ($\epsilon e$) which is model dependent and may vary depending on its origin and model parameters.
 For each models one can treat $ \epsilon$ as a free parameter being agnostic about its origin and underlying assumptions.
However, to show how the results differ in presence of a tiny induced coupling we pick a benchmark value $\epsilon=-\frac{g_X}{70}$. This particular benchmark has already been used in some previous works on light $Z'$ models and we have taken the same benchmark to facilitate easy comparison of our constraints with other works \cite{Banerjee:2021laz,Escudero:2019gzq}.
%
In the following section we show the results for all models (with $\mathbb{X}_1=0$) for both $\epsilon=0$ and $\epsilon=-g_X/70$, and the qualitative change for other values of $\epsilon$ can be easily deduced from our analysis.
Needless to say that with the increase in $M_{Z'}$, the energy density of $Z'$ gets suppressed and hence $N_{\rm eff}$ decreases for both of the above mentioned two cases as shown in Appendix \ref{sec:mzvary}.

The excess in $N_{\rm eff}$ at CMB in the presence of $Z'$ can be used to constrain the parameter space, as the deviation from SM predicted value of $N_{\rm eff}$ is also very precisely constrained from Planck 2018 \cite{Planck:2018vyg}.
We display the constraints on all popular $U(1)_X$ models by performing exhaustive numerical scans on the parameter space (for $N_{\rm eff}$), in the next section.

\section{Specific $U(1)_X$ Symmetries}
\label{sec:examples}

 In this section, we will discuss some phenomenological consequences of various well motivated $U(1)_X$ symmetries.
As mentioned earlier, while doing this we consider that the corresponding gauge boson $Z'$ was initially in thermal equilibrium and the RHNs are heavy enough that they do not take part in $\nu_L$ decoupling.
 In order to estimate $N_{\rm eff}$, we find the temperature ratio $T_{\nu_L}/T_\gamma$ at $T_\gamma=T_{\rm CMB}$ by solving eq.\eqref{eq:nux}-\eqref{eq:Tgamma} and then plug into eq.\eqref{eq:neff}.
As described earlier the value of $N_{\rm eff}$ is strongly dependent on parameters: $g_X$ and $M_{Z'}$, and hence the upper limit on $N_{\rm eff}$ by Planck 2018 \cite{Planck:2018vyg} leads to strong constraint on the same parameter space 
\footnote{Note that, strictly speaking the Planck 2018 limit on $N_{\rm eff}=2.99^{+0.34}_{-0.33}$ is obtained from fitting Planck 2018 data assuming only SM particle content \cite{Planck:2018vyg}. 
However, just like several previous works on this topic \cite{EscuderoAbenza:2020cmq,Escudero:2018mvt,Escudero:2019gzq,Esseili:2023ldf,Li:2023puz}, we simply portray this bound to constrain light BSM  mediator. 
A complete refitting of the Planck 2018 data considering the new particle as well can potentially shift the central value and confidence levels a bit\cite{Sabti:2019mhn}, though
 a dedicated analysis of refitting is beyond the scope of this paper.}.
Before going into the details of the results we summarize the existing laboratory-based constraints on light $Z'$ in Table-\ref{tab:cons}. 
The detailed discussion on existing constraints can be found in Ref.  \cite{Bauer:2018onh,Coloma:2020gfv,Coloma:2022umy,AtzoriCorona:2022moj,DeRomeri:2024dbv,Majumdar:2021vdw,Demirci:2023tui}.
In addition to the constraints mentioned in the table, the light $Z^\prime$ may also encounter astrophysical constraints like SN1987A \cite{Fiorillo:2022cdq,Fiorillo:2023ytr,Janka:2017vlw,Akita:2023iwq}. However, this supernova constraint differs depending on the specific gauge model and requires a detailed analysis which is beyond the scope of this work.
\begin{table}[h]
\begin{center}
\begin{tabular}{| c|c|c|c|c|}
  \hline
Experiments        & Process  &  Observations  & Refs.     \\
\hline \hline
\makecell{E$\nu$ES \\ (elastic electron-neutrino \\ scattering) } &  \makecell{$\nu  ~e\to \nu~ e$ \\ incoming particle:\\ solar $\nu$ $(\nu_e)$}& recoil rate of $e^-$  &  \makecell{ XENON \cite{Majumdar:2021vdw,A:2022acy}, \\ LZ\cite{DeRomeri:2024dbv},\\  Borexino \\ \cite{Coloma:2022umy,Khan:2019jvr,AtzoriCorona:2022moj} } \\
\hline
\makecell{CE$\nu$NS \\ (coherent elastic neutrino\\-nucleus scattering) } &  \makecell{$\nu  ~N\to \nu~ N $ \\ nuclei N: \{CsI, Ar\}}& recoil rate of nucleus  &  \makecell{ CsI and Ar \\ \cite{Cadeddu:2020nbr,COHERENT:2021xmm,Banerjee:2021laz} \\
XENON \cite{Majumdar:2021vdw,DeRomeri:2024dbv}}\\
\hline
\makecell{Fixed target experiments } &  \makecell{$e(p) ~N \to e (p) ~N~ Z^\prime$ \\ $Z^\prime \to e^+ e^-$ }& \makecell{displaced vertex \\ with di-electron}  &   \makecell{ E137 \cite{Bjorken:1988as}, \\  E141 \cite{Riordan:1987aw},\\  E774 \cite{Bross:1989mp} etc.} \\
\hline
\makecell{Neutrino trident } &  \makecell{$\nu~N \to \nu~N~\mu^+~\mu^-$ }& \makecell{ di-muon final state } &   \makecell{LBNE \cite{Altmannshofer:2014pba}, \\ CCFR \cite{CCFR:1991lpl},\\ DUNE \cite{Altmannshofer:2019zhy}} \\
\hline
\makecell{ATLAS and CMS } &  \makecell{$(i)~ p p \to Z^\prime \to \mu^+\mu^-$\\ $(ii)~ p p \to h, \phi \to Z^{*} Z^\prime $, \\ $ Z^*~Z^\prime \to 4 \ell $}& \makecell{ $(i)$ oppositely \\charged muons \\ $(ii)$ $4\ell$ final state}  &   \makecell{$(i)$ \cite{CMS:2019buh} \\ $(ii)$ \cite{ATLAS:2014jlg, CMS:2012bw}} \\
\hline
\makecell{ BaBar, Belle } &  \makecell{$e^+~e^- \to \gamma_{\rm ISR} ~Z^\prime$ \\, $Z^\prime \to \ell^+ \ell^- (\ell=e,\mu)$ }& \makecell{ lepton charged\\ tracks with $\gamma$}  &   \makecell{BaBar \cite{BaBar:2014zli,Bauer:2018onh}\\ Belle \cite{Inguglia:2016acz}} \\
\hline
\makecell{ KLOE } & \makecell{$e^+~e^- \to \gamma_{\rm ISR} Z^\prime$,\\ $Z^\prime \to \mu^+ \mu^- , \pi^+ \pi^-$  } & \makecell{ charged tracks\\ with $\gamma$} &    \makecell{KLOE \cite{KLOE-2:2018kqf}} \\
\hline
  \end{tabular}
\end{center}
\caption{Summary of the existing laboratory-based constraints on light $Z^\prime$. For further details see  Ref.  \cite{Bauer:2018onh,Coloma:2020gfv,Coloma:2022umy,AtzoriCorona:2022moj,DeRomeri:2024dbv,Majumdar:2021vdw}.}
  \label{tab:cons}
\end{table}

For the ease of readers, we denote the experimental constraints with the same notation for all $U(1)_X$
models shown in Fig.\ref{fig:b_l}-\ref{fig:b1_3lmu}.
The bounds from E$\nu$ES and CE$\nu$NS from direct detection experiments (XENON1T) are shown by dark cyan and dark blue solid lines respectively.
On the other hand bounds from Borexino and coherent scattering from CsI$+$Ar material are depicted by gray hatched region and dark green solid line repectively.
The magenta solid line and the light blue hatched region in some figures signify the constraint from $\nu$ oscillation data \cite{Coloma:2020gfv} and fixed target experiments respectively.
Finally, the $2\sigma$ upper limit on $N_{\rm eff}$ at CMB ($3.33$) from Planck 2018 \cite{Planck:2018vyg} is shown by the red dashed line. To display the variation of $N_{\rm eff}$ we also show two additional contours for  $N_{\rm eff}=3.2$ and $N_{\rm eff}=3.5$ depicted by blue dashed lines.
We now discuss each of the $U(1)_X$ models one by one throughout this section.
Although  two of them have already been discussed in the literature \cite{Escudero:2019gzq,Esseili:2023ldf}, 
for completeness and comparative analysis, here we discuss all  popular $U(1)_X$ models (see Table. \ref{tab:all}).

\subsection{The Gauged $U(1)_{B-L}$ Symmetry}
\label{sec:B-L}
\begin{figure}[!tbh]
\centering
\subfigure[\label{bl}]{
\includegraphics[scale=0.4]{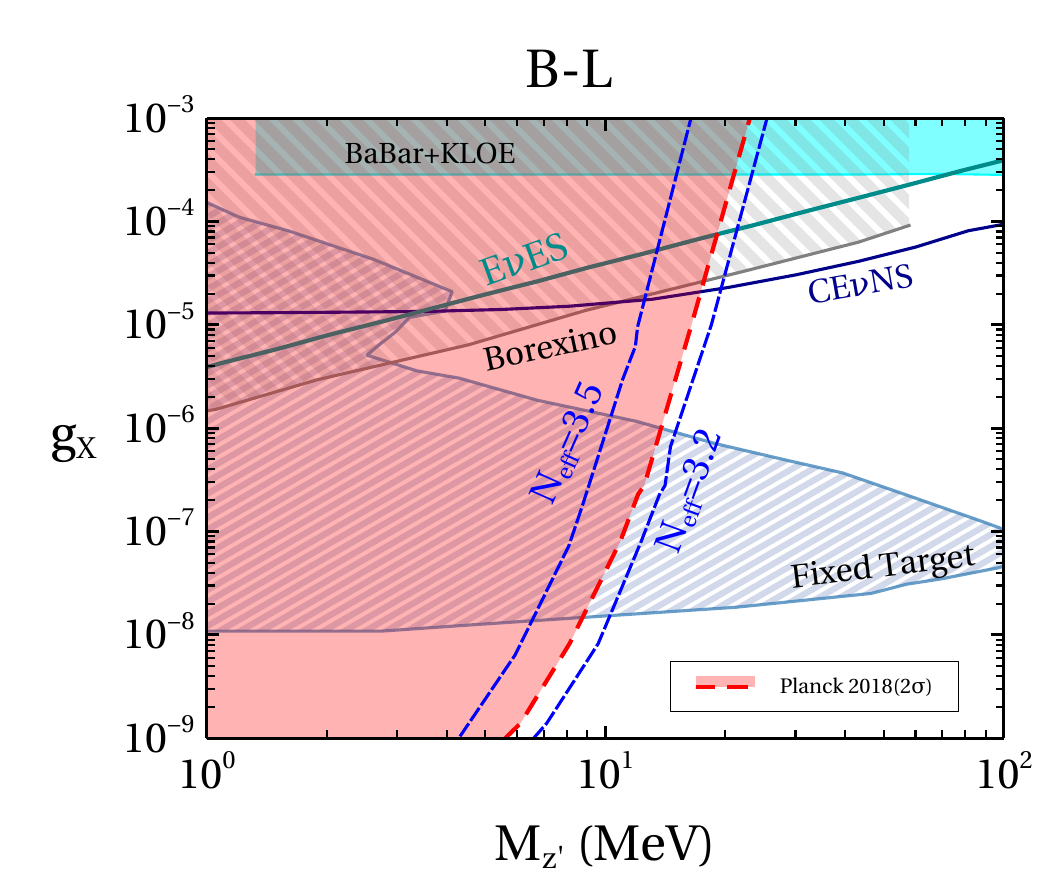}}
\subfigure[\label{ble}]{
\includegraphics[scale=0.4]{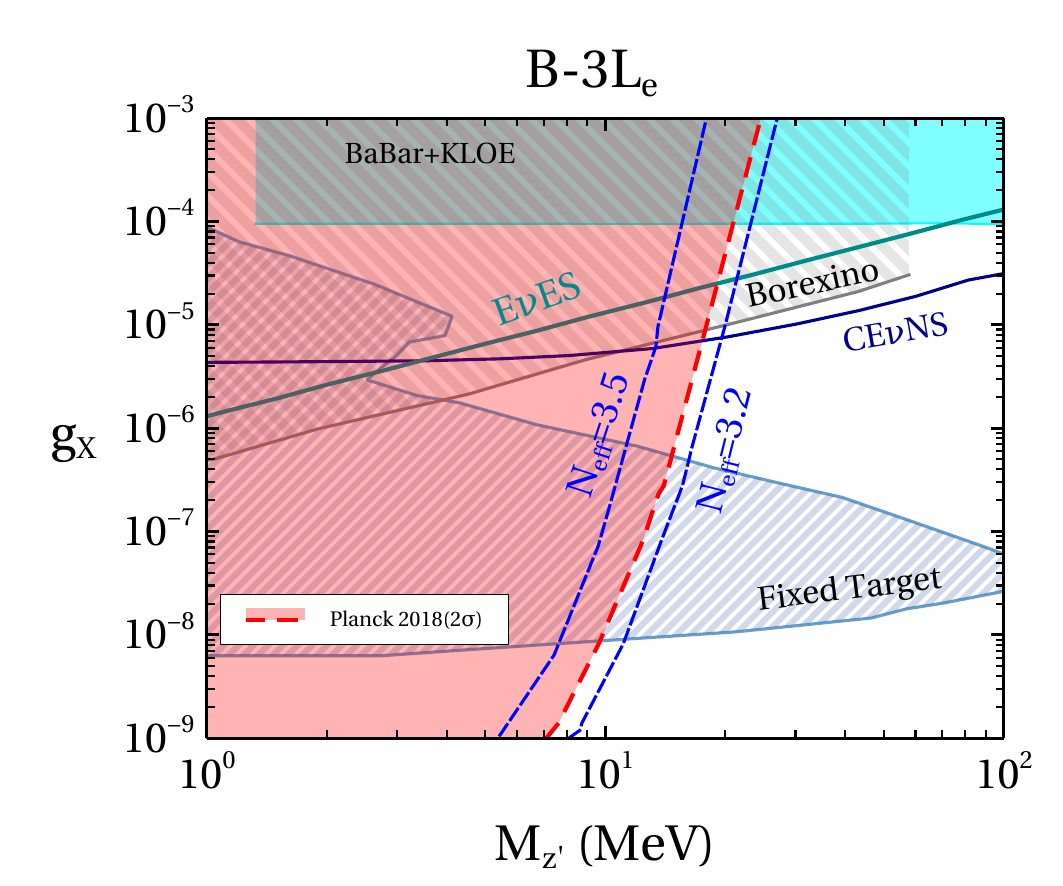}}
\subfigure[\label{blmu_nl}]{
\includegraphics[scale=0.4]{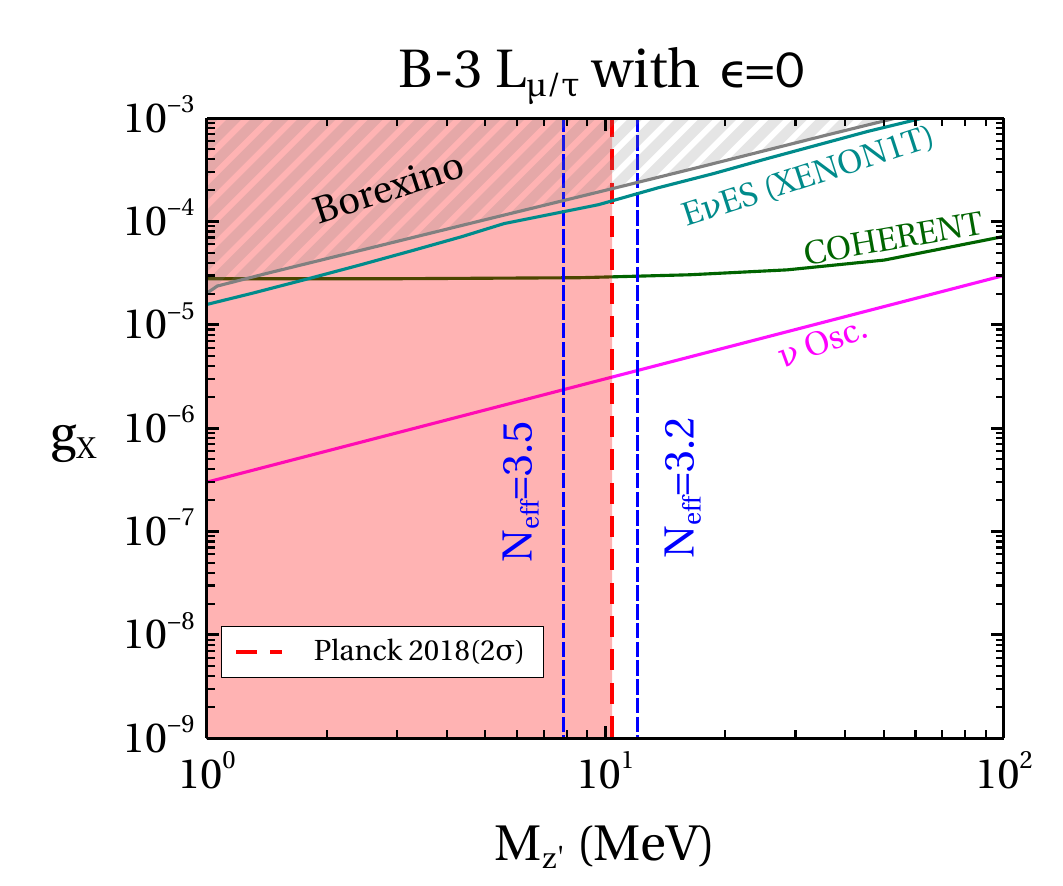}}
\subfigure[\label{blmu_loop}]{
\includegraphics[scale=0.4]{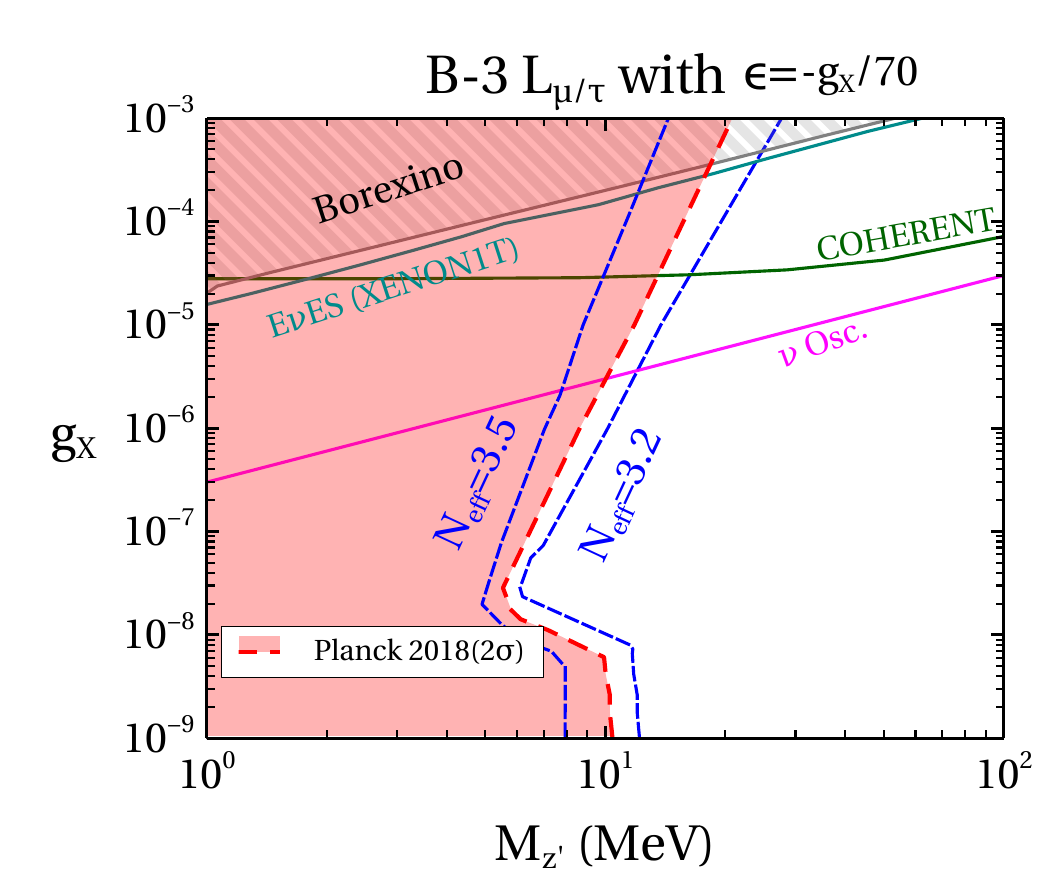}}
\caption{Constraints from $N_{\rm eff}$ (Planck 2018) along with experimental bounds shown in the parameter space of light $Z'$ realized in (a) $B-L$, (b) $B -3 L_e$, (c) $B - 3L_{\mu}~\bl{(B - 3L_{\tau})}$ with $\epsilon=0$ and (d) $B-3 L_\mu~\bl{(B - 3L_{\tau})}$ with $\epsilon=-g_X/70$. The $2\sigma$ upper limit from Planck 2018 on $N_{\rm eff}$ is shown by the red dashed lines and the exclusion region is shown by the red shaded area. } 
\label{fig:b_l}
\end{figure}

$B-L$ type of $U(1)_X$  model has been studied exhaustively in different contexts \cite{Ma:2014qra,Ma:2015mjd,Majumdar:2021vdw,Mandal:2023oyh}.
In such gauge extensions, all three generation of leptons are charged with $U(1)_X$ charge $|\mathbb{X}_i|=1$.
Hence it lies to the first class of models which have tree level coupling of $Z'$ with $e^\pm$ as discussed in Sec.\ref{sec:neff}. 
We show various constraints in $M_{Z'}$ vs. $g_X$ plane in Fig.\ref{bl}.
Since, in this model $Z'$ can couple to both leptons and quarks it attracts strong constraints from fixed target experiments in the sub-GeV range of $M_{Z'}$ with $g_X\gtrsim 10^{-8}$ \cite{Riordan:1987aw, Bjorken:1988as,Bross:1989mp}.
A summary of these bounds can be found in Refs. \cite{Bjorken:2009mm,Harnik:2012ni}.
On the other hand, neutrino scattering experiments like E$\nu$ES and CE$\nu$NS from XENON1T, XENONnT and LZ \cite{Majumdar:2021vdw,A:2022acy,DeRomeri:2024dbv}  place stringent constraint in the parameter space.
Neutrino electron scattering from  Borexino also excludes $g_X\gtrsim10^{-5}$ for $M_{Z'}\sim 10$ MeV \cite{Coloma:2022umy,Khan:2019jvr,AtzoriCorona:2022moj}. 
Other experiments like BaBar \cite{BaBar:2014zli,BaBar:2017tiz} and KLOE \cite{KLOE-2:2018kqf} can place constraint at relatively larger couplings ($g_X\sim 10^{-4}$). 
In spite of these stringent constraints, there is significant parameter space, particularly for low mass and small couplings which remains unconstrained. 
The constraint coming from the upper limit on $N_{\rm eff}$ from Planck 2018 \cite{Planck:2018vyg},  shown by the red dashed line, can exclude a large portion of this parameter space as shown in Fig.\ref{bl}.
It excludes $M_{Z'}\lesssim 5$ MeV at lower couplings ($g_X\sim 10^{-9}$) and $M_{Z'}\lesssim20$ MeV at higher couplings ($g_X\sim 10^{-3}$).
As discussed in Sec.\ref{sec:neff} with an increase in $g_X$ the relevant collision terms increase leading to an enhancement in $N_{\rm eff}$.
And with an increase in $M_{Z'}$ the effect in $N_{\rm eff}$ decreases due to the suppressed energy density of $Z'$ at $\nu_L$ decoupling. 
For this reason, we observe the contour for the upper limit on $N_{\rm eff}$ gradually shifts towards the higher $M_{Z'}$ with increase in $g_X$.
This phenomenon can also be understood from the contour lines (blue dashed) with $N_{\rm eff}$ values of $3.5$ and $3.2$.

\subsection{The Gauged $U(1)_{B-3L_i}$ Symmetry}
\label{sec:B-3Li}
So far we have considered only the flavour universal $U(1)_X$ symmetry where each generation of quarks and charged leptons carry the same charge under the $U(1)_X$ symmetry. However, there are several flavour dependent $U(1)_X$ symmetries known in the literature \cite{Bonilla:2017lsq,Coloma:2020gfv,Coloma:2022umy,AtzoriCorona:2022moj,DeRomeri:2023ytt}. 
One of such flavour dependent $U(1)_X$ symmetry  is the so called  $U(1)_{B-3L_i}$ gauge symmetries \cite{Ma:1997nq}. 
In this scenario, the $U(1)_X$ charges of all generations of quarks are still the same as that we consider in $B-L$ symmetry,
however only one generation of SM leptons carry the $U(1)_X$ charge. 
There are three types for such symmetries namely the $U(1)_{B-3L_e}$, $U(1)_{B-3L_\mu}$ and $U(1)_{B-3L_\tau}$ gauge symmetries 
corresponding to the first, second, and third generation of SM leptons respectively. 
The charges of the SM particles and new particles under these symmetries are given in Table~\ref{tab:all}.

\subsubsection{\bf{${B-3L_e}$ model:}}
In the $B-3 L_e$ model the light $Z'$ couples to $e^\pm$ apart from all 3 quark generations \cite{DeRomeri:2023ytt}.
As the $Z'$ has tree level coupling with $e^\pm$ it will lie in the first class of
models in deciding $N_{\rm eff}$ as discussed in Sec.\ref{sec:neff}.
Hence, the upper bound on $N_{\rm eff}$ arising from Planck 2018 also constrains this model, 
as depicted by the tilted red dashed line (similar to $B-L$) shown in Fig.\ref{ble}.
However, in this case $U(1)_X$ charge of $e^\pm$, $|\mathbb{X}_1|=3$ so the constraints will be stronger than $B-L$ model. 
As the coupling of $Z'$ with $e^\pm$ is thrice ($\mathbb{X}_1 g_X$) than that of $B-L$ model,
the relevant BSM processes also become larger compared to $B-L$, leading to larger $N_{\rm eff}$ for the same $g_X$ and $M_{Z'}$.
For this reason we notice the contour for the upper limit on $N_{\rm eff}$ shift towards the right (to higher $M_{Z'}$) as compared to $B-L$ shown in Fig.\ref{ble}.
For the same reason the constraints like E$\nu$ES and CE$\nu$NS from Xenon1T \cite{Majumdar:2021vdw}, Borexino \cite{Coloma:2022umy,Khan:2019jvr,AtzoriCorona:2022moj} become more stringent compared to $B-L$.
Similar argument holds for  BaBar \cite{BaBar:2014zli,BaBar:2017tiz}, KLOE \cite{KLOE-2:2018kqf} and fixed target experiments \cite{Bross:1989mp,Bauer:2018onh,AtzoriCorona:2022moj}.

\subsubsection{${B-3L_{\mu/\tau}}$ {\bf model:}}

In contrast to the $B-3 L_e$ model, in ${B-3L_\mu}$ (${B-3L_\tau}$) model the light $Z'$ couples to only second (third) generation of leptons with charge $|\mathbb{X}_2|=3$ ($|\mathbb{X}_3|=3$).
As the $Z'$ has no tree level coupling with $e^\pm$ it will lie in the second class of
models in deciding $N_{\rm eff}$ as discussed in Sec.\ref{sec:neff}.
If we do not assume any induced coupling between $Z'$ and $e^\pm$, then the only BSM process to affect $N_{\rm eff}$ is the decay of $Z'$ to $\nu_L$ sector with $100\%$ branching ratio (BR). 
Thus all the $Z'$ density eventually dilutes to $\nu_L$ sector independent of the coupling ($X_{2/3}g_X$) and hence we observe the bound from $N_{\rm eff}$ becomes a straight line in the $M_Z'$ vs. $g_X$ plane as shown in Fig.\ref{blmu_nl}.
However, the situation changes drastically if we consider some induced coupling 
{\footnote{we pick this benchmark value to compare with the existing literature \cite{Banerjee:2021laz}}}
$\epsilon=-g_X/70$ between $Z'$ and $e^\pm$
and the corresponding $N_{\rm eff}$ constraint is portrayed in 
Fig.\ref{blmu_loop}.
The knee like pattern is due to the interplay between decay and scattering and is discussed in great detail in our earlier work \cite{Ghosh:2023ilw}.
In the presence of the induced coupling 
both the processes (i) scattering: $e^+e^-\to \nu \Bar{\nu}$ mediated by $Z'$ and (ii) decay $Z'\to e^+e^-/ \nu \Bar{\nu}$
are active in the higher coupling ($g_X\gtrsim 10^{-8}$) regime and the effect on $N_{\rm eff}$ becomes almost similar like $B-3L_e$ in this regime.
However, in the lower coupling regime $g_X\lesssim 10^{-8}$ the scatterings are suppressed ($\sim (g_X/70)^2$) than the decay ($\sim g_X^2$), and hence the BSM process becomes decay dominated \cite{Ghosh:2023ilw}. Thus in this regime ($g_X\lesssim 10^{-8}$) the constraint from $N_{\rm eff}$ becomes independent of the coupling $g_X$.
In the absence of tree level coupling with $e^\pm$,
 the experimental constraints in the parameter space are comparatively less stringent than those for $B-3L_e$. 
 The major constraints come from
E$\nu$ES from Xenon1T \cite{Majumdar:2021vdw}, Borexino \cite{Coloma:2022umy,Khan:2019jvr,AtzoriCorona:2022moj} and COHERENT data \cite{AtzoriCorona:2022moj}.
Additionally, $\nu$ oscillation data also constrains the parameter space as given in  Ref. \cite{Coloma:2020gfv}.

\subsection{The Gauged $U(1)_{L_i - L_j}$ Symmetry}
\label{sec:Li-Lj}
So far we have been discussing about the $U(1)_X$ gauge symmetries under which 
at least some of the quarks were always charged.
However, anomaly free models can be constructed by assigning $U(1)_X$ charges to even lepton sectors only.
Such $U(1)_X$ models are $L_\mu - L_{\tau},~L_e - L_{\mu},$ and $L_e - L_{\tau}$ models. We start our discussion $L_\mu - L_{\tau}$ type models due to its wide discussion in the existing literature  \cite{Escudero:2019gzq,Banerjee:2021laz}.  

\begin{figure}[!tbh]
\centering
\subfigure[\label{w1}]{
\includegraphics[scale=0.4]{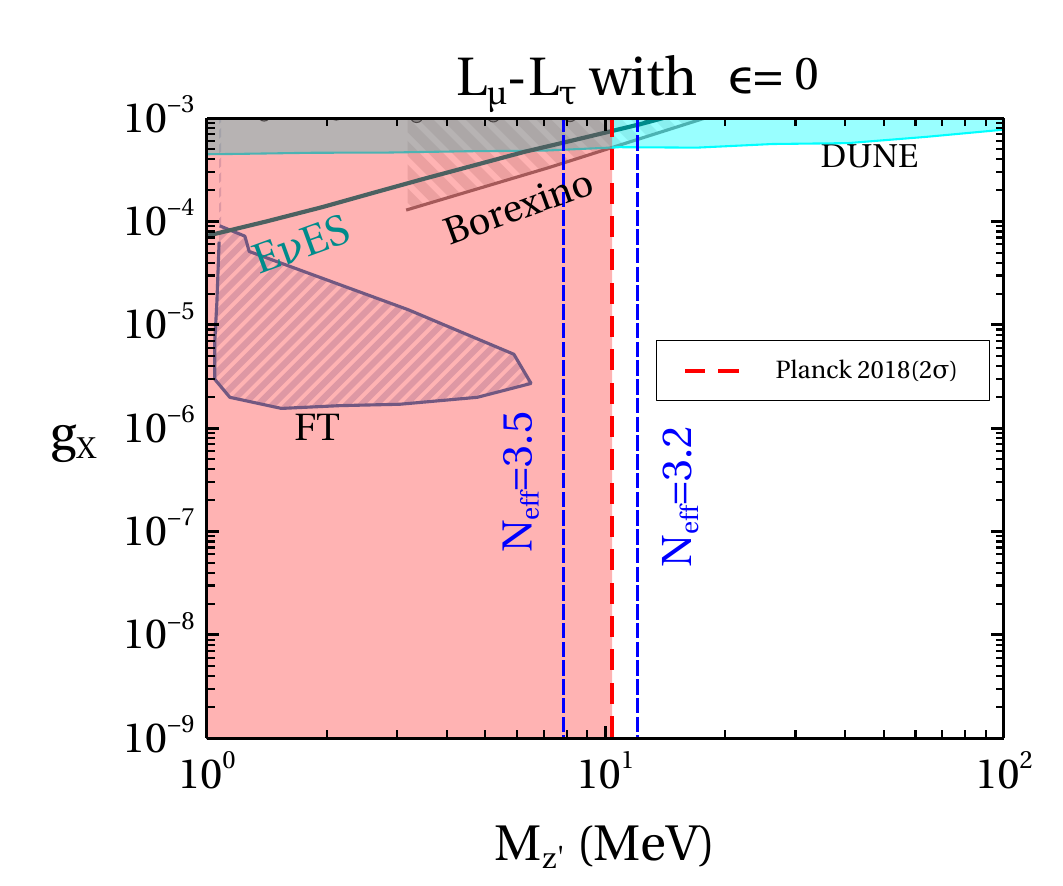}}
\subfigure[\label{w2}]{
\includegraphics[scale=0.4]{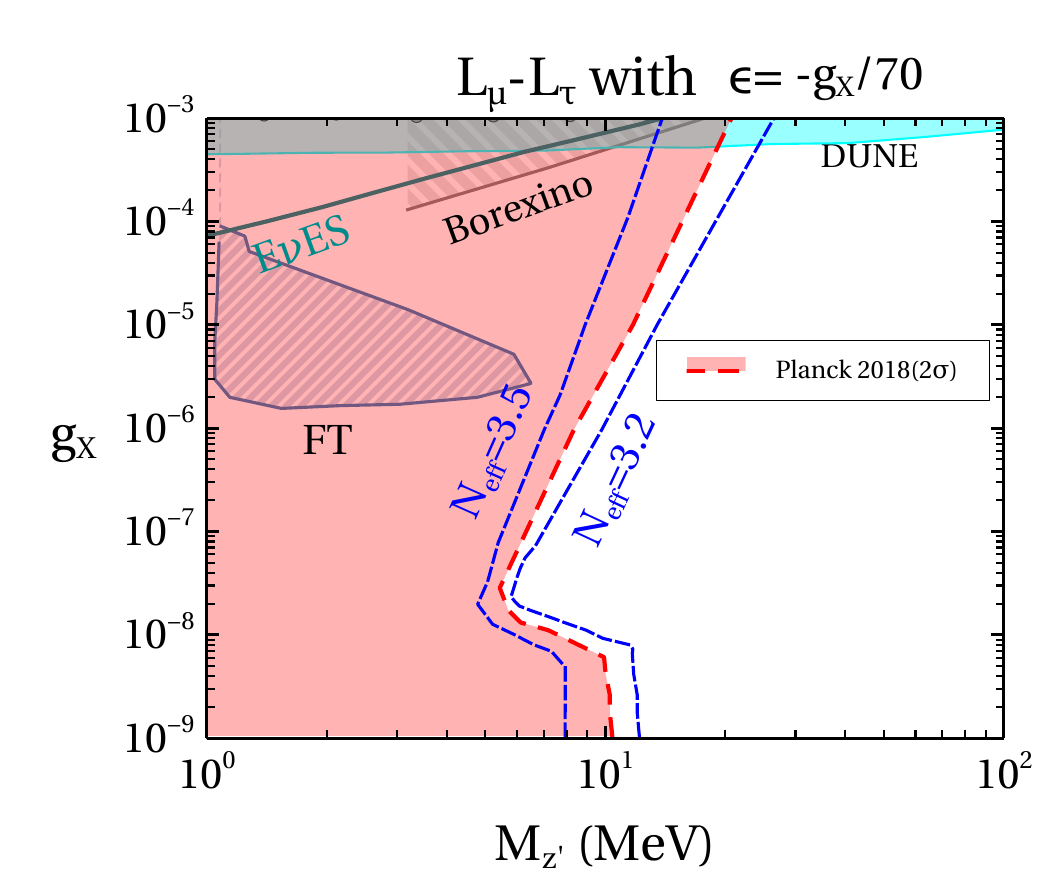}}
\subfigure[\label{lemu}]{
\includegraphics[scale=0.4]{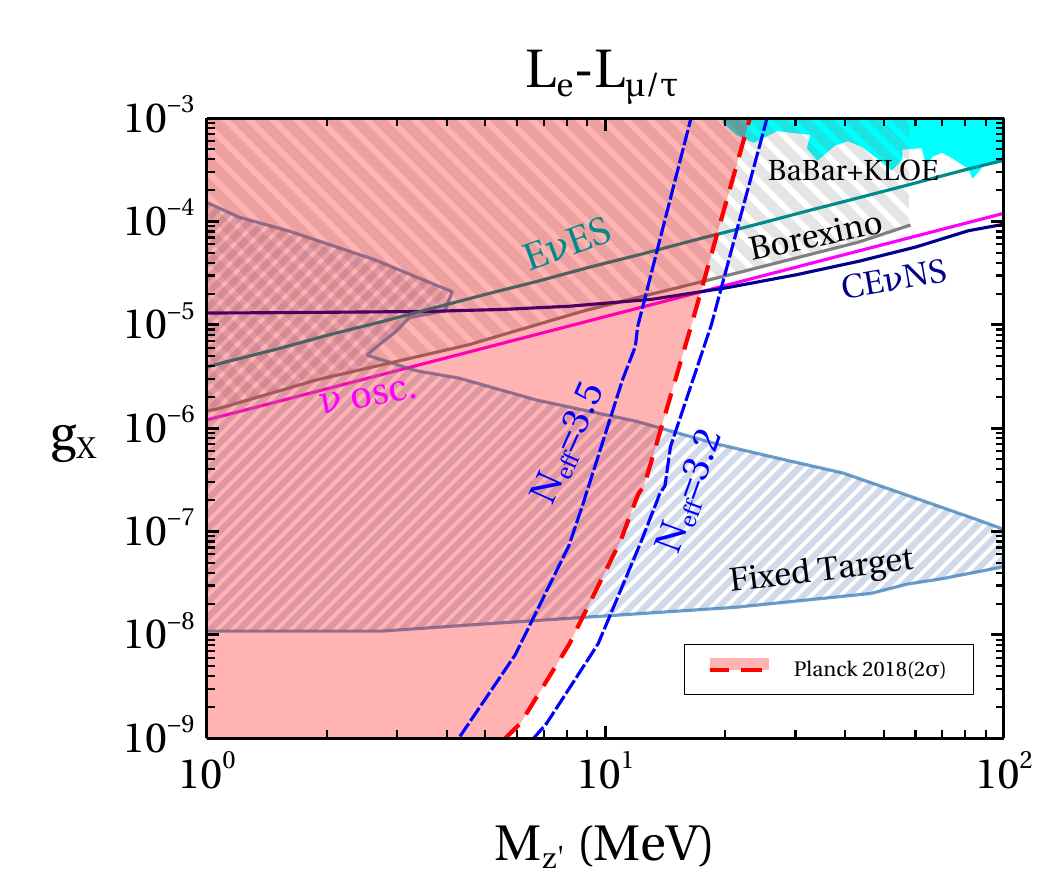}}
\caption{Constraints from $N_{\rm eff}$ (Planck 2018) along with experimental bounds shown in the parameter space of light $Z'$ realized in  (a) $L_{\mu}-L_\tau$ with $\epsilon=0$, (b) $L_{\mu}-L_\tau$ with $\epsilon=-g_X/70$ and (c) $L_e-L_\mu~\bl{(L_e-L_\tau)}$. The $2\sigma$ upper limit from Planck 2018 on $N_{\rm eff}$ is shown by the red dashed lines and the exclusion region is shown by red shaded area. } 
\label{fig:li_lj}
\end{figure}

\subsubsection{${L_\mu - L_{\tau}}$ \bf{model:}}
The $N_{\rm eff}$ upper limit contour in $L_\mu - L_{\tau}$ symmetry also exhibits similar dependence likewise in the $B-3 L_\mu$ model, due to the absence of tree level $Z'e^+e^-$ vertex. The constraints from $N_{\rm eff}$ with  and  without the induced couplings are shown in 
Fig.\ref{w1} and Fig.\ref{w2} respectively.
In the sub-GeV range the other relevant constraints are E$\nu$ES from Xenon1T \cite{Majumdar:2021vdw}, Borexino \cite{Coloma:2022umy,Khan:2019jvr,AtzoriCorona:2022moj}.
Fixed target (FT) experiment has very loose constraints due to the absence of both $Z'$ quark coupling and $Z'$ electron coupling \cite{DeRomeri:2024dbv}. 


\subsubsection{${L_e - L_{\mu/\tau}}$ \bf{model:}}
In $L_e - L_\mu$ (or $L_e - L_\tau$ ) model tree level $Z'e^+e^-$ vertex exists in contrast to $L_\mu - L_{\tau}$ type model and hence the constraint from $N_{\rm eff}$ is exactly same as $B-L$ model as depicted in Fig.\ref{lemu}.
For the same reason constraints from E$\nu$ES from Xenon1T \cite{DeRomeri:2024dbv}, Borexino \cite{Coloma:2022umy,Khan:2019jvr,AtzoriCorona:2022moj}, Babar \cite{BaBar:2014zli,Bauer:2018onh} and
fixed target experiments \cite{Bauer:2018onh,Bross:1989mp} are found to be  stronger than  $L_\mu - L_\tau$ model. 

\subsection{The Gauged $U(1)_{B_i-3L_j}$ Symmetry}
\label{sec:Bi-3Lj}

There are also $U(1)_X$ symmetries where the symmetry is flavour dependent in both quarks as well as the lepton sector namely the $U(1)_{B_i-3L_j}$ gauge symmetries~\cite{Bonilla:2017lsq}. 
In this class, there are total nine such symmetries: $B_1-3 L_j,~B_2-3 L_j,~ B_3-3 L_j~(j\equiv e,\mu,\tau)$, where $B_1, B_2,B_3$ correspond to the first, second and third generation of quarks respectively.
For all nine cases, the charges of the $\nu_{R_i}$ can be fixed by the anomaly cancellation conditions.
Among these nine symmetries, we discussed about $B_3-3 L_j$ symmetry in our previous work \cite{Ghosh:2023ilw}.
In this work, we will shed light on the remaining six symmetries. The charge assignments of these symmetries are shown in Table~\ref{tab:all}.


\subsubsection{$\mathbf{B_i-3 L_e}$ \bf{model:}}
\begin{figure}
    \centering
    \includegraphics[scale=0.4]{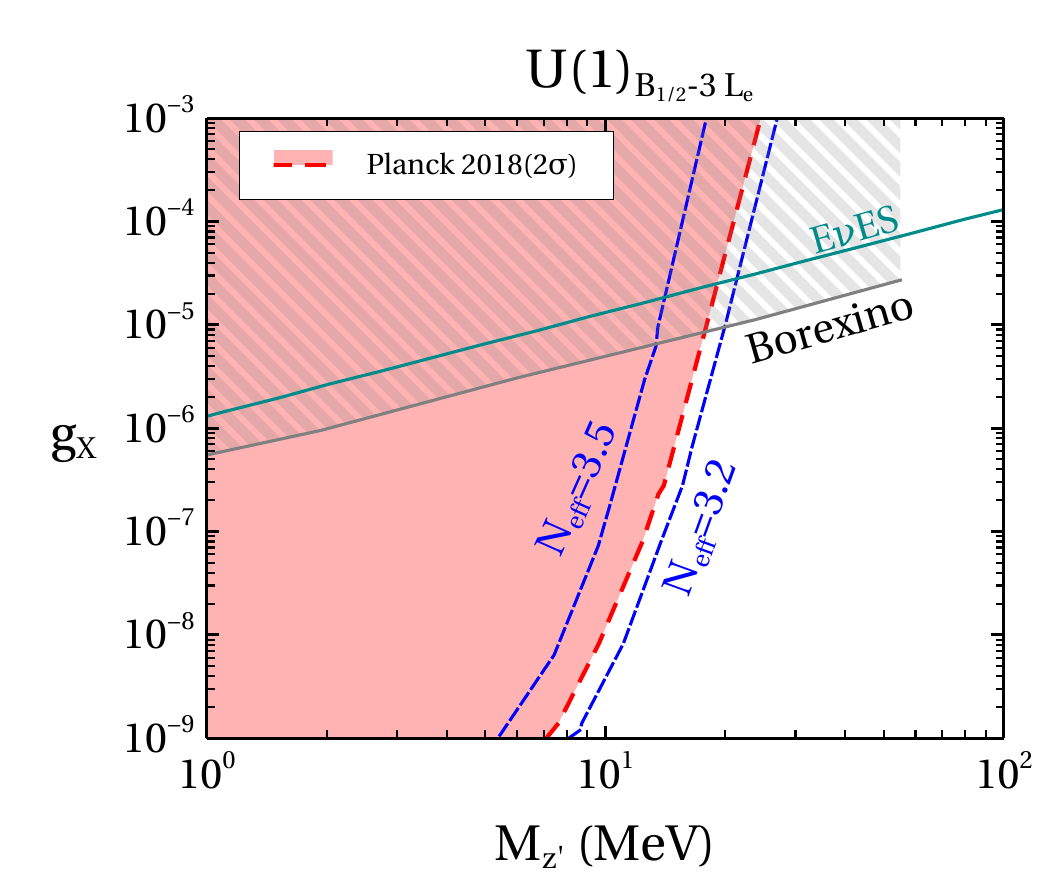}
    \caption{Constraints from $N_{\rm eff}$ (Planck 2018) along with experimental bounds shown in the parameter space of light $Z'$ realized in $B_{1/2}-3 L_e$. The $2\sigma$ upper limit from Planck 2018 on $N_{\rm eff}$ is shown by the red dashed lines and the exclusion region is shown by the red shaded area. }
    \label{fig:b13le}
\end{figure}

Models like $B_1-3 L_e,~ B_2-3 L_e,~B_3-3 L_e$ fall in this category where $Z'$ has tree level coupling with $e^\pm$ and any one generation of quarks.
Due to the presence of tree level $Z'e^+e^-$ vertex (with $|\mathbb{X}_1|=3$) the $N_{\rm eff}$ increases with increase in $g_X$ for a fixed $M_{Z'}$. Hence the contour for the upper limit on $N_{\rm eff}$  exhibits a pattern similar to the $B-3 L_e$ model. 
The constraints are shown in Fig.\ref{fig:b13le}.
For comparison, we also showcase bounds from E$\nu$ES (from Xenon1T) \cite{Majumdar:2021vdw}, Borexino \cite{Khan:2019jvr} which are similar to the case in $B-3 L_e$ model, since these bound rely only on the lepton coupling.
Note that for $B_{1}-3 L_e$ type of model with the tree level coupling of $Z'$ to first generation quarks (and hence nucleons) may attract constraints from CE$\nu$NS, BaBar, and fixed target experiments. 
However such analysis has not been yet performed with these types of models in the existing literature and dedicated analysis is beyond the scope of this work.
For $B_3- 3 L_e$ model due to the presence of $Z'-b$ coupling bound from $B_s-\Bar{B_s}$ mixing is also applicable as shown in Ref. \cite{Ghosh:2023ilw}.

\subsubsection{$\mathbf{B_i-3 L_{\mu/\tau}}$ \bf{model:}}
\begin{figure}
    \centering
    \subfigure[\label{b1lmu_1}]{
    \includegraphics[scale=0.4]{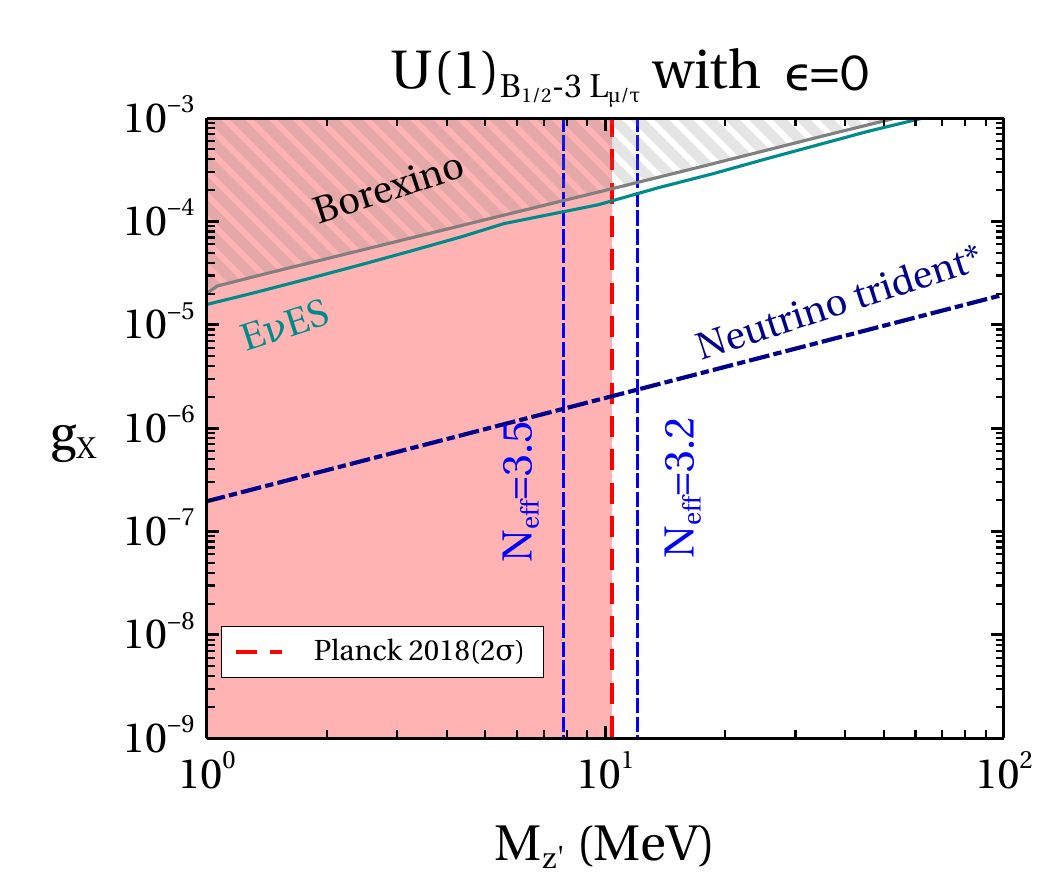}}
     \subfigure[\label{b1lmu_2}]{
    \includegraphics[scale=0.4]{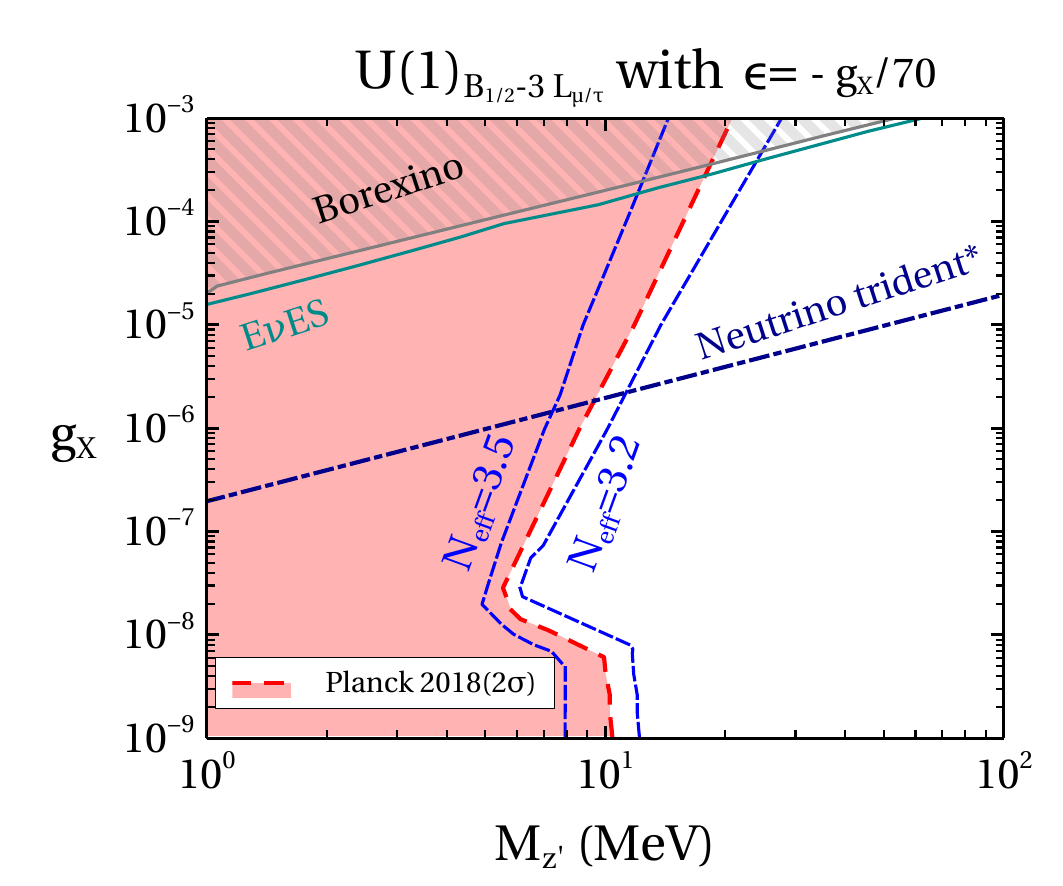}}
    \caption{Constraints from $N_{\rm eff}$ (Planck 2018) along with experimental bounds shown in the parameter space of light $Z'$ realized in  (a) $B_{1/2}-3 L_{\mu}~\bl{(B_{1/2}-3 L_{\tau})}$ with $\epsilon=0$, (b) $B_{1/2}-3 L_{\mu}~\bl{(B_{1/2}-3 L_{\tau})}$ with $\epsilon=-g_X/70$. The $2\sigma$ upper limit from Planck 2018 on $N_{\rm eff}$ is shown by the red dashed lines and the exclusion region is shown by the red shaded area.
    The bound on $B_{1/2}-3 L_\tau$ parameter space will be similar to $B_{1/2}-3 L_\mu$ model except 
    neutrino trident bound which is applicable to $B_1-3 L_\mu$ type model only.}
    \label{fig:enter-label}
    \label{fig:b1_3lmu}
\end{figure}

In contrast to the previous case, in $B_i-3 L_{\mu/\tau}$ model $Z'$ has no tree level coupling with $e^\pm$ and thus the $N_{\rm eff}$ contour becomes independent of the coupling $g_X$ as shown in Fig.\ref{b1lmu_1}.
However, in presence of induced coupling $\epsilon=-g_X/70$ the contour for $N_{\rm eff}$ follow pattern similar to $B-3 L_{\mu/\tau}$ as shown in Fig.\ref{b1lmu_2}. 
The knee like pattern of the contour for the upper bound on $N_{\rm eff}$ is due to the interplay between scattering and decay as already discussed earlier.
Similar to the earlier case $B_i-3 L_{\mu/\tau}$ attracts constraints like  E$\nu$ES from Xenon1T \cite{Majumdar:2021vdw}, Borexino \cite{Khan:2019jvr,AtzoriCorona:2022moj} which are similar to the case in $B-3 L_\mu$ model.
Here also we do not portray the bounds that may come from nuclear interactions.
Additionally, $B_i-3 L_{\mu}$ type of model attracts strong constrain from neutrino trident experiment \cite{Bonilla:2017lsq}.

Thus the precise measurement of $N_{\rm eff}$ by Planck 2018 data \cite{Planck:2018vyg} provides a unique opportunity to probe light $Z'$ gauge bosons realized in various $U(1)_X$ models.
In general, most of the terrestrial experiments looking for light $Z'$ lose sensitivity in the sub-GeV mass range of $Z'$. 
Although the neutrino scattering experiments can constrain $M_{Z'}\ll\mathcal{O}(1)$ MeV, it can not probe  very low couplings ($g_X \lesssim 10^{-6}$).
On the other hand, the upper limit on $N_{\rm eff}$ by Planck 2018 can constrain $M_{Z'}\lesssim\mathcal{O}(1)$ MeV as well as very small couplings even upto $g_X\sim 10^{-9}$, as shown in this section. 
The value of $N_{\rm eff}$ relies on the $\nu_L$ decoupling and in the SM scenario it takes place around $T\sim $MeV \cite{Escudero:2018mvt}.
Hence $N_{\rm eff}$ is sensitive to BSM particles with sufficient energy at that temperature as in our case $Z'$ with mass $M_{Z'}\sim 10$ MeV. 
For the same reason $N_{\rm eff}$ becomes insensitive  to slightly heavier $Z'$ with $M_{Z'}\gtrsim30$ MeV,
when the energy density becomes Boltzmann suppressed. 
Nevertheless the upper limit on $N_{\rm eff}$ can probe the parameter space which is beyond the reach of current ground based experiments.
Astrophysical bounds like supernova may constrain very small couplings \cite{Akita:2022etk}, however, such bounds largely depend on the modeling of the stellar objects \cite{Bar:2019ifz}. 
Whereas the bound from $N_{\rm eff}$ is relatively more generic since here the only assumption is $Z'$ to be a thermal bath which is the case for $g_X\gtrsim 10^{-9}$ \cite{Escudero:2019gzq}.


\section{Conclusions}
\label{sec:conclusions}
The stringent limit on $N_{\rm eff}$ by Planck 2018 \cite{Planck:2018vyg} can be used to constrain various BSM paradigms involving light ($\sim \mathcal{O}$(MeV)) mediators.
In this work, we show that the $\nu_L$ decoupling is significantly affected in the presence of light $Z'$ particles interacting with $\nu$ (or, $e^\pm$) leading to an enhanced $N_{\rm eff}$.
This in turn leads to constraining the parameter space of light $Z'$ in the mass ($M_{Z'}$) vs. coupling ($g_X$) plane using the Planck 2018 data.
As the SM $\nu_L$ decoupling usually takes place at $T\sim 1$ MeV, $N_{\rm eff}$ upper limit puts strong constraint for $M_{Z'}\lesssim$ few MeV.
However, the ground based experiments looking for light BSM particles lose sensitivity in the sub-MeV range, and thus the constraint from $N_{\rm eff}$ can be useful in this regard.
We consider several popular $U(1)_X$ models like $B-L, ~B-3 L_i, ~L_i-L_j,~ B_i-3 L_i$ and study the contribution in $N_{\rm eff}$ due to the light $Z'$ realized in this model assuming $Z'$ was initially in thermal bath.
In our previous work \cite{Ghosh:2023ilw}, we showed the signature in $N_{\rm eff}$ due to the light $Z'$ can be broadly classified into two categories depending on whether  $Z'$ has tree level coupling with $e^\pm$ or not.
We correctly identify the aforementioned $Z'$ models and display the constraints from $N_{\rm eff}$ along with other existing constraints.
For models with tree level $Z'e^+e^-$  coupling (e.g. $B-L,~B-3 L_e,~L_e-L_{\mu/\tau},~B_i-3L_e $), the constraint from $N_{\rm eff}$ becomes stronger with an increase in $g_X$ because of the enhancement in BSM processes with higher values of $g_X$.
In the absence of tree level $Z'e^+e^-$  coupling $N_{\rm eff}$ is dominated by the decay, $Z'\to \nu \bar{\nu}$ and hence becomes insensitive to $g_X$.
Thus we observe the $N_{\rm eff}$ constraints to be straight line in $M_{Z'}$ vs. $g_X$ plane for such models (e.g. $B-3 L_{\mu/\tau},~L_\mu-L_{\tau},~B_i-3L_{\mu/\tau} $).
However, in the presence of induced $Z'e^+e^-$ couplings, the constraints modify due to the interplay of decay scattering as discussed in Sec.\ref{sec:examples}.
We observe that in a significant parameter space, specifically for low $Z'$ mass and small couplings, the constraint from $N_{\rm eff}$ is much more stringent than the experimental searches. In fact in many cases, the $N_{\rm eff}$ provides the only constraint on such parameter space.
Thus looking at the footprints in $N_{\rm eff}$
at CMB can be an useful scheme to probe various light BSM particles having interactions with $e^\pm$ or, $\nu_L$. 


\begin{acknowledgments}
SJ is funded by CSIR, Government of India, under the NET JRF fellowship scheme with Award file No. 09/080(1172)/2020-EMR-I. PG acknowledges the financial support
through the APEX project at the Institute of Physics, Bhubaneswar. PG also acknowledges local hospitality at ICTS Bengaluru during the visit for the SATPP 2024.
\end{acknowledgments}

\appendix
\section{Variation of $N_{\rm eff}$ for different  $U(1)_X$ models }
\label{app:neff_all}
\begin{figure}[!tbh]
    \centering
    \includegraphics[scale=0.5]{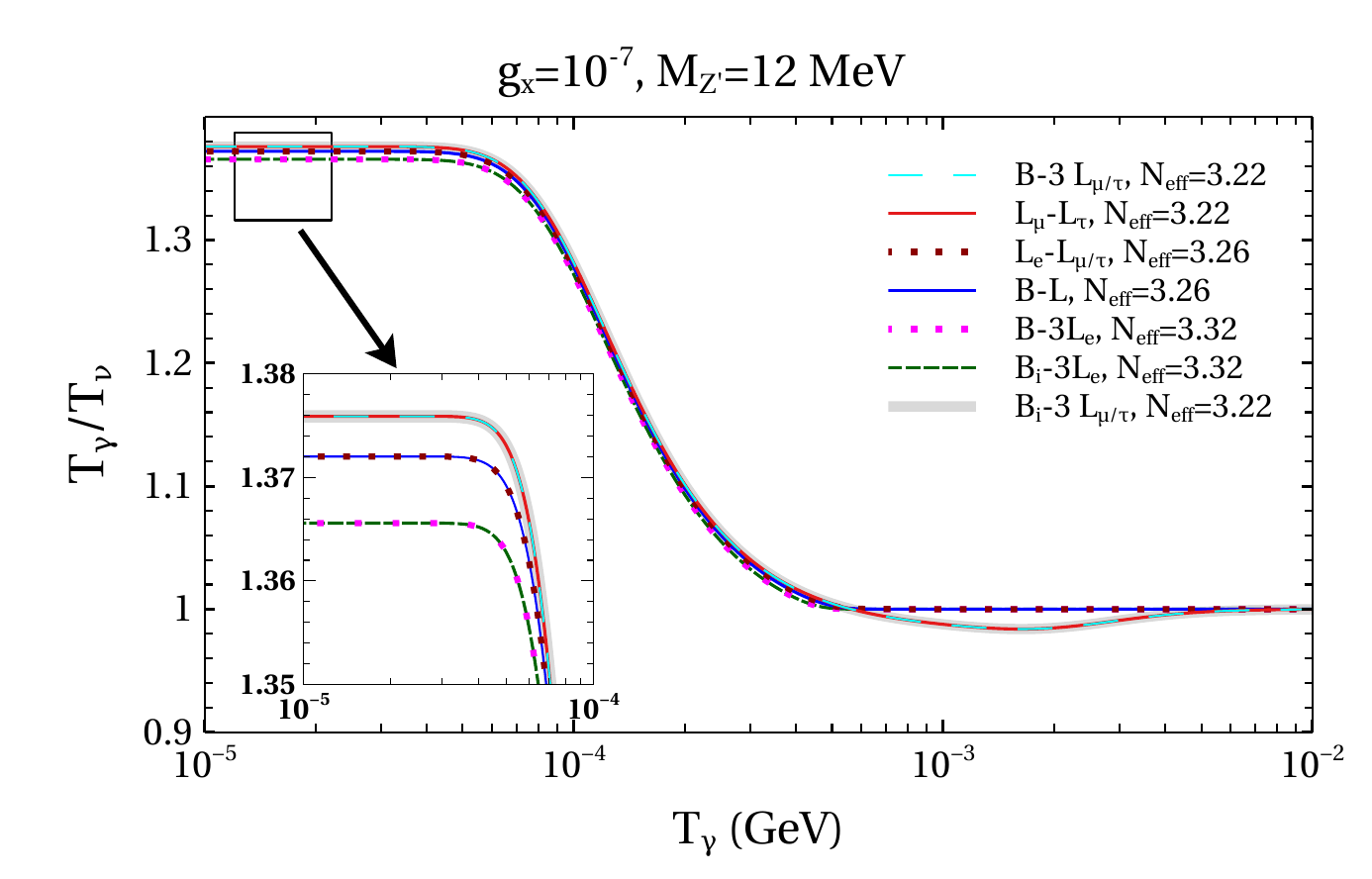}
    \caption{Variation of the temperature ratio $T_{\gamma}/T_{\nu}$ with photon bath temperature $T_{\gamma}$  for $U(1)_X$ gauge boson mass $M_{Z'}=12$ MeV and gauge coupling $g_X=10^{-7}$. The different colored lines correspond to different $U(1)_X$ models and are depicted by the plot legends in the figure.
    Note that here we assumed all 3 $\nu_{L}$ share same temperature. The low temperature behaviour of the lines are zoomed in and shown in the plot in inset.}
    \label{fig:tnu_all}
\end{figure}

Here we discuss the variation of $N_{\rm eff}$ for different $U(1)_X$ models. To evaluate $N_{\rm eff}$ one needs to find the temperature ratio $T_\nu/T_\gamma$ at low temperature ($T_{\rm CMB}$) as mentioned in eq.\eqref{eq:neff}. 
For computing the temperature ratio we simultaneously solve the coupled Boltzmann equations in eq.\eqref{eq:nux}-\eqref{eq:Tgamma}. 
As elaborated earlier in sec.\ref{sec:neff} the  BSM contribution encoded in the energy transfer rates between photon and neutrino bath in the temperature equations depends mainly on the BSM interaction of $\nu_i$ and $e^\pm$.
A detailed formulation of the collision terms in presence of generic $U(1)_X$ gauge boson has been made in Appendix (B) of ref.\cite{Ghosh:2023ilw}.
Adopting the formalism from there we solve the aforementioned temperature equations and display our numerical results in the plane $T_\gamma$ vs. $T_\gamma/T_\nu$ in Fig.\ref{fig:tnu_all}.

In Fig.\ref{fig:tnu_all} we choose a specific benchmark parameter value $g_X=10^{-7}$, $M_{Z'}=12$ MeV. 
The temperature ratio curves corresponding to different $U(1)_X$ models are depicted by different coloured lines: $B-3 L_{\mu/\tau}$ (cyan dashed), $L_{\mu}-L_{\tau}$ (red solid), $L_{e}-L_{\mu/\tau}$ (dark red dotted), $B-L$ (blue solid), $B-3 L_{e}$ (magenta dotted), $B_i-3 L_{e}$ (dark green dashed), $B_i-3 L_{\mu/\tau}$ (thick gray solid).
At high temperature ($T_\gamma \sim 10$ MeV) all the curves merge to $T_\gamma/T_\nu=1.0$ as at that temperature the $\nu_L$ and $\gamma$ bath has not decoupled yet.
However, at $T_\gamma \lesssim 0.5$ MeV $e^\pm$ becomes non-relativistic transferring its entropy to $\gamma$ bath and hence increases the temperature $T_\gamma$. 
For this reason we observe a gradual rise in the  $T_\gamma/T_\nu$ curves after the temperature drops down $T_\gamma \lesssim 0.5$ MeV.
The temperature ratio becomes constant at very low temperature ($T_\gamma\lesssim 3\times 10^{-4}$) as all the relevant interaction becomes suppressed after that temperature.

The box in the inset left corner of the Fig.\ref{fig:tnu_all} contains a zoomed in version of the temperature ratio curves for $10^{-5}{\rm ~MeV}\le T_\gamma \le 10^{-4}$ MeV. Note that all the $U(1)_X$ models lead to $N_{\rm eff}$
different than that of SM predicted value \cite{Escudero:2018mvt}. The different values of $N_{\rm eff}$ for different $U(1)_X$ models in the plot can be easily understood from the fact that the only relevant BSM interactions for $\nu_L$ decoupling are between $e^\pm$, $\nu_i$, $Z'$ as argued  in sec.\ref{sec:neff}. We highlight the key points of the aforementioned figure below.

\begin{itemize}
    \item {\bf Models with $\mathbb{X}_1=0$:} For models like $L_\mu-L_\tau$, $B-3 L_{\mu/\tau}$, $B_i-3 L_{\mu/\tau}$ we observe the same value $N_{\rm eff}=3.22$. For such models without tree level $Z'e^+e^-$ vertex the only BSM contribution is from the decay $Z'\to \nu_i\Bar{\nu_i}$. Hence for any values of $Z'\nu_{i}\Bar{\nu_{i}}$ coupling ($\equiv \mathbb{X}_{2,3} g_X$) all the $Z'$ density eventually dilutes to $\nu_L$ bath and thus enhance $N_{\rm eff}$. For that reason despite the change in $X_2$ and $X_3$ these models exhibit same value of $N_{\rm eff}$.

    \item {\bf Models with $\mathbb{X}_1\neq0$:} For models with $\mathbb{X}_1=1$ like $L_e-L_{\mu/\tau}$ and $B-L$
    we note the same value of $N_{\rm eff}=3.26$. 
    For this models both the decays $Z'\to \nu_i\Bar{\nu_i}$ , $Z'\to e^+e^-$ and $Z'$ mediated scattering $e^+e^-\to \nu_i\Bar{\nu_i}$ are  relevant. In this case, since BSM contribution is mostly scattering dominated, governed by the effective coupling $\mathbb{X}_1 g_X$, despite the different values of $\mathbb{X}_2, \mathbb{X}_3$ these models exhibit similar imprints in $N_{\rm eff}$.

    However, for models with $\mathbb{X}_1=3$ like
    $B-3 L_{e}$, $B_i-3 L_{e}$,
    the effecting coupling $\mathbb{X}_1 g_X$ increases leading to an enhancement in the BSM contribution.
    For this reason we observe higher values of $N_{\rm eff}=3.32$ than that of previously mentioned models.
\end{itemize}

Thus the imprints  of $N_{\rm eff}$ in presence of such $U(1)_X$ gauge boson can be distinctly classified as done in sec.\ref{sec:neff}.

\section{Variation in $N_{\rm eff}$ with $M_{Z'}$}
\label{sec:mzvary}
\begin{figure}[!tbh]
    \centering
    \includegraphics[scale=0.5]{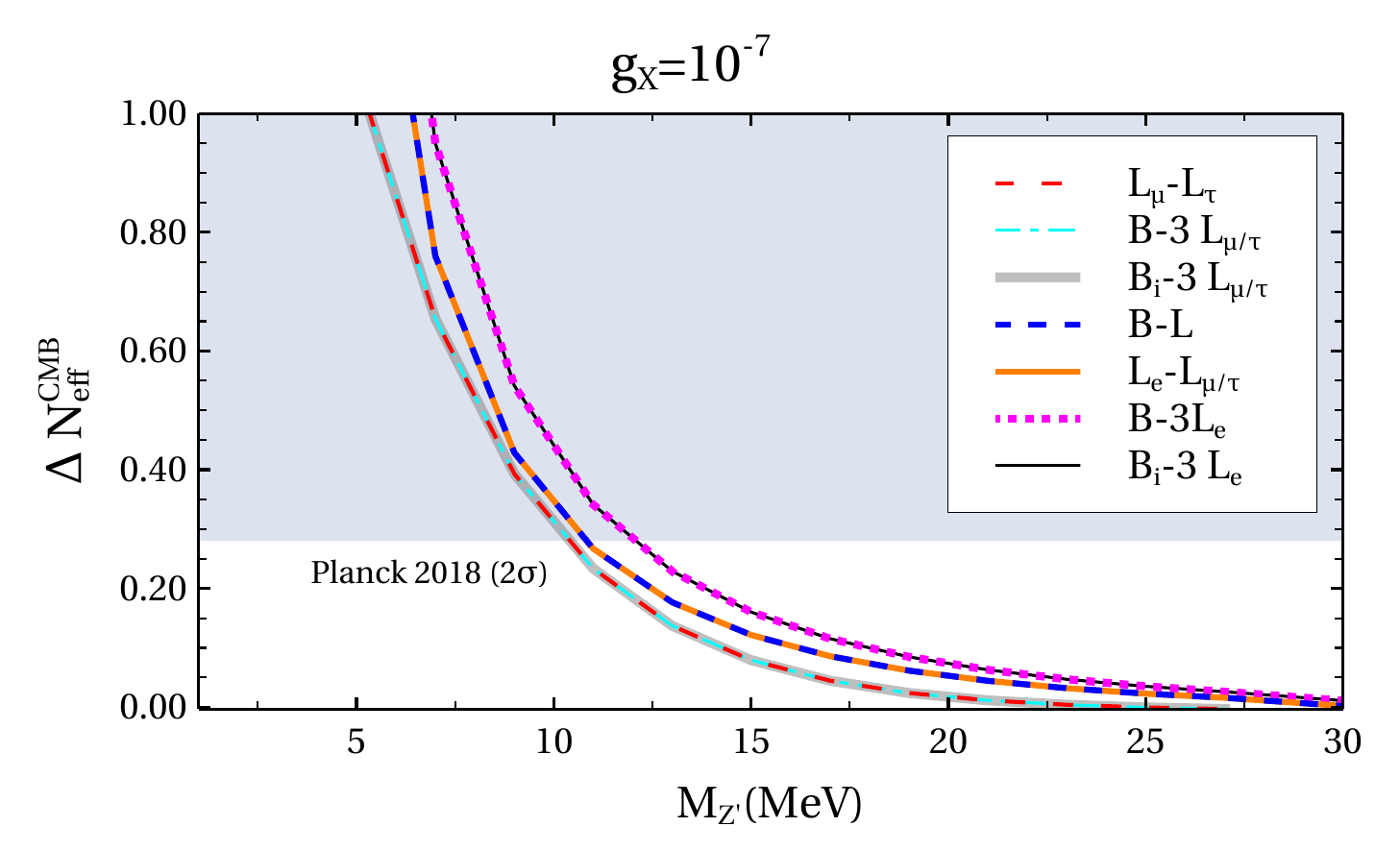}
    \caption{Variation of $\Delta N_{\rm eff}\equiv N_{\rm eff}-N_{\rm eff}^{\rm SM}$ with gauge boson mass $M_{Z'}$ for all popular $U(1)_X$ models for a fixed coupling $g_X=10^{-7}$. The lines corresponding  to different $U(1)_X$ models are presented by different color and styles shown in the plot legends. }
    \label{fig:mz_all}
\end{figure}

In Fig.\ref{fig:mz_all} we portray the variation in $N_{\rm eff}$ with the gauge boson mass $M_{Z'}$ for different $U(1)_X$
models. To amplify the change in $N_{\rm eff}$ for this figure we plot $\Delta N_{\rm eff}\equiv N_{\rm eff}-N_{\rm eff}^{\rm SM}$ in the $y-$axis.
The curves corresponding to different $U(1)_X$ models are depicted by different coloured lines:  $L_{\mu}-L_{\tau}$ (red dashed), $B-3 L_{\mu/\tau}$ (cyan dot dashed), $B_i-3 L_{\mu/\tau}$ (thick gray solid), $B-L$ (blue dashed),  $L_{e}-L_{\mu/\tau}$ (orange solid), $B-3 L_{e}$ (magenta dotted), $B_i-3 L_{e}$ (dark green solid).
For all models we observe the decrease in $\Delta N_{\rm eff}$
with an increase in $M_{Z'}$. As with higher mass the $Z'$ density becomes suppressed the BSM contribution also gets suppressed leading to less value of $\Delta N_{\rm eff}$.
As in the SM scenario the $\nu_L$ decoupling takes place at $\sim 1$ MeV \cite{Dodelson:2003ft}, for $M_{Z'}\gtrsim 30$ MeV
the extra $Z'$ hardly plays any role in deciding $N_{\rm eff}$.

For fixed $M_{Z'}$ we note that $3$ class of models \\
(i)$L_{\mu}-L_{\tau}$ (red dashed), $B-3 L_{\mu/\tau}$ (cyan dot dashed), $B_i-3 L_{\mu/\tau}$ (thick gray solid); \\
(ii)  $B-L$ (blue dashed),  $L_{e}-L_{\mu/\tau}$ (orange solid); and \\
(iii) $B-3 L_{e}$ (magenta dotted), $B_i-3 L_{e}$ (dark green solid) \\
lead to 3 distinct values  of $\Delta N_{\rm eff}$ for the reason already discussed in the previous section. 

\bibliographystyle{utphys}
\bibliography{bibliography}

\end{document}